%% file: main.tex
\documentclass[journal]{IEEEtran}
\usepackage{graphicx}
\usepackage{overpic}
\usepackage{amsmath}
\usepackage{amssymb}
\usepackage{multirow}
\usepackage{makecell}
\usepackage{array}
\usepackage{stfloats}
\usepackage{gensymb}
\usepackage[switch]{lineno}
\usepackage{float}
\usepackage{booktabs}
\usepackage{color}
\usepackage{cite}
\usepackage[implicit=false]{hyperref}
\usepackage{soul}
\usepackage{pifont}
\usepackage{lineno}
\usepackage{epsfig}
\usepackage{textcomp}
\usepackage{threeparttable}
\usepackage{makecell}
\usepackage{color}
\usepackage{xcolor,colortbl}
\usepackage{enumerate}
\usepackage{ulem}
\usepackage{amsmath}
\usepackage{algorithm}
\usepackage{algpseudocode}
\usepackage[numbers]{natbib}
\begin{document}
% \linenumbers
\normalem
\title{Multi-modal Uncertainty Robust Tree Cover Segmentation For High-Resolution \\Remote Sensing Images}
\author{Yuanyuan Gui, Wei Li, ~\IEEEmembership{Senior Member,~IEEE}, Yinjian Wang, Xiang-Gen Xia ~\IEEEmembership{Fellow,~IEEE}, Mauro Marty, Christian Ginzler, Zuyuan Wang

\thanks{This paper is supported by the National Key R\&D Program of China (Grant No.2021YFB3900502) and was carried out in the framework of the Swiss National Forest Inventory (NFI). (Corresponding author: Wei Li, Zuyuan Wang).}

\thanks{Y. Gui,  W. Li and Y. Wang are with the School of Information and Electronics, Beijing Institute of Technology,  and the National Key Laboratory of Science and Technology on Space-Born Intelligent Information Processing, Beijing 100081, China, and also with  Beijing Institute of Technology, Zhuhai, Guangdong 519088, China (e-mail: 3220205108@bit.edu.cn, liwei089@ieee.org, yinjw@bit.edu.cn).}

\thanks{X.-G. Xia is with Department of Electrical and Computer Engineering, University of Delaware, Newark, DE 19716, USA (e-mail: xxia@ece.udel.edu).}

\thanks{M. Marty, C. Ginzler and Z. Wang are with Swiss Federal Institute for Forest, Snow, and Landscape Research WSL, CH-8903, Birmensdorf, Switzerland (email: mauro.marty@wsl.ch, christian.ginzler@wsl.ch, zuyuan.wang@wsl.ch).}
}
% The paper headers
\markboth{}%
{Shell}

\maketitle
\begin{abstract}
Recent advances in semantic segmentation of multi-modal remote sensing images have significantly improved the accuracy of tree cover mapping, supporting applications in urban planning, forest monitoring, and ecological assessment. Integrating data from multiple modalities—such as optical imagery, light detection and ranging (LiDAR), and synthetic aperture radar (SAR)—has shown superior performance over single-modality methods. However, these data are often acquired days or even months apart, during which various changes may occur, such as vegetation disturbances (e.g., logging, and wildfires) and variations in imaging quality. Such temporal misalignments introduce cross-modal uncertainty, especially in high-resolution imagery, which can severely degrade segmentation accuracy. To address this challenge, we propose MURTreeFormer, a novel multi-modal segmentation framework that mitigates and leverages aleatoric uncertainty for robust tree cover mapping. MURTreeFormer treats one modality as primary and others as auxiliary, explicitly modeling patch-level uncertainty in the auxiliary modalities via a probabilistic latent representation. Uncertain patches are identified and reconstructed from the primary modality’s distribution through a VAE-based resampling mechanism, producing enhanced auxiliary features for fusion. In the decoder, a gradient magnitude attention (GMA) module and a lightweight refinement head (RH) are further integrated to guide attention toward tree-like structures and to preserve fine-grained spatial details. Extensive experiments on multi-modal datasets from Shanghai and Zurich demonstrate that MURTreeFormer significantly improves segmentation performance and effectively reduces the impact of temporally induced aleatoric uncertainty.
\end{abstract}


\IEEEpeerreviewmaketitle

\section{Introduction}
\label{sec:intro}
\input{1introduction}

\section{Related Works}
\label{sec:relatedworks}
\input{2Relatedwork}

\section{Methodology}
\label{sec:methodology}
\input{3Methodology}

\section{Experiments}
\label{sec:experiments}
\input{4Experiments}

\section{Conclusions}
\label{sec:conclusion}
\input{5conclusion}

\bibliographystyle{IEEEtranN}
\bibliography{gyy.bib}

\end{document}

%% file: 1Introduction.tex
With billions of hectares of tree cover worldwide~\cite{hansen2013high}, the function of trees is critical for ecology and biodiversity~\cite{dwyer1991significance,schweingruber1996tree}. Transparent high-accuracy tree cover monitoring is therefore vital for large-scale carbon capture~\cite{mo2023integrated}, urban afforestation and reforestation management~\cite{dwyer1991significance}. Traditional tree mapping methods, e.g., manual delineation, have been changed to automated deep learning-based approaches~\cite{li2016deep} because of the rapid development of remote sensing interpretation technologies. Among these methods, semantic segmentation has high interpretation efficiency and accuracy due to its end-to-end design. It fulfills and improves the requirements of rapid tree cover mapping~\cite{gao2021fine,gui2022infrared}.

Currently, three primary modalities of remote sensing images are used for tree cover mapping, namely optical remote sensing (ORS)~\cite{prasad2011optical}, synthetic aperture radar (SAR)~\cite{moreira2013tutorial}, and light detection and ranging (LiDAR)~\cite{reutebuch2005light}. Each modality has its own distinct advantages and limitations and all the three modalities have been widely applied to tree cover mapping tasks~\cite{brandt2023wall, magnard2016single, pearse2018comparison}. ORS provides rich spectral information and high spatial resolution, but its performance can be affected by weather and illumination conditions. SAR enables all-weather, day-and-night imaging with strong structural penetration but limited spectral diversity. LiDAR captures precise 3D structural details of forest canopies and is often used to generate digital surface models (DSMs) with excellent canopy delineation capabilities. However, its application is restricted by limited coverage and high acquisition costs~\cite{jin2022fusion}.

In order to utilize the advantages of each modality more effectively, researches use multi-modal remote sensing images for tree cover mapping, and in particular, combining ORS with SAR or ORS with LiDAR, which has demonstrated superior performance compared to single-modal segmentation~\cite{Shao8039160}.~\citet{lehmann2015sar} investigated the interoperability of ORS and SAR data for the purpose of large-scale and operational forest monitoring using the canonical variate analysis method.~\citet{dong2013mapping} accurately mapped rubber plantations in tropical regions by integrating SAR with multi-temporal Landsat observations of distinctive phenological characteristics using the decision tree method. These approaches utilize the complementarity between ORS and SAR to extract enhanced information of various environmental variables, including land cover, above-ground biomass, road networks, and crop types. In terms of the combination of LiDAR and ORS,~\citet{Paris6835211} developed a three-dimensional model-based approach for the estimation of the tree top height based on both ORS and LiDAR data for precise individual tree crown detection and delineation.~\citet{liu2017mapping} successfully surveyed the urban tree species in Surrey by combining LiDAR-derived canopy structural information with hyperspectral vegetation indices through a Random Forest classifier. While all these studies demonstrate that multi-modal data fusion outperforms single-source methods in tree classification and segmentation, none of the studies has provided insight into how temporal variations in multi-modal inputs affect high-resolution tree cover mapping.%~\citet{qin2022individual} used the watershed-spectral-texture-controlled normalized cut (WST-Ncut) algorithm to delineate individual trees in a subtropical broadleaf forest after combining the LiDAR with ORS data. 

Despite significant advances in remote sensing technologies in terms of spatial resolution and data availability, acquiring synchronous multi-source observations in the same area remains challenging~\cite{liu2024review,liu2024multimodal}. For example, the LiDAR-based SwissALTI3D digital terrain model, derived from airborne laser scanning (ALS) and designed to exclude vegetation and man-made structures, captures Swiss bare-earth topography and undergoes systematic updates on a six-year cycle~\cite{swisstopo_sar}. High-resolution multispectral imagery requires a three-year cycle to achieve complete coverage of Switzerland~\cite{swisstopo_rgb}. The revisit period of the ORS-equipped GF-7 satellite in the same area is about 5 to 7 days, while that of the SAR-equipped GF-3 satellite is about 3 to 4 days~\cite{chen2022introduction}, and the effective data acquisition may be delayed due to weather conditions, solar illumination, and cloud cover. This temporal mismatch introduces significant cross-modal uncertainty in tree cover mapping tasks so that frequent vegetation changes including pruning, harvesting, or seasonal defoliation may reduce semantic segmentation accuracy. Existing tree cover mapping studies predominantly assume that the multi-modal data are temporally aligned.

In this paper, we propose a semantic segmentation framework named MURTreeFormer to deal with uncertainty and maximally preserve the complementary information. Specifically, it addresses the challenges of aleatoric uncertainty in multi-modal remote sensing imagery and enhances the accuracy of tree cover mapping. The multi-modal inputs are considered as one primary and one auxiliary modalities, and patch-level uncertainty is estimated through a probabilistic modeling approach. A distributional feature representation for the auxiliary modality is reconstructed based on the latent distribution of the primary modality using a VAE-based mechanism. The proposed selective uncertainty-guided reconstruction module
(SURM) mitigates cross-modal inconsistencies and improves the robustness of multi-modal feature fusion. Furthermore, three task-specific modules have been integrated into MURTreeFormer to enhance its representation capabilities. The cross-modal distillation module (CDM) is embedded in the encoder to align heterogeneous modality features; the gradient magnitude attention (GMA) module is incorporated into the decoder to guide attention using luminance gradients; and a lightweight shallow feature refinement head (RH) is applied at the final stage to preserve fine-grained spatial details of tree crowns.

In summary, the main contributions of this work are as follows:
\begin{itemize}
    \item To model patch-level uncertainty, enabling fine-grained identification of ambiguous regions that hinder cross-modal feature alignment.
    
    \item To introduce a novel VAE-based patch reconstruction strategy, in which auxiliary modality features are generated from the primary modality’s latent distribution, thereby improving alignment and reducing uncertainty.
    
    \item To propose a high-performance multi-modal semantic segmentation network for fine-grained tree-cover mapping. By integrating multiple task-oriented modules, the model achieves superior performance on two challenging datasets.
    
    \item To offer broader ecological significance in future applications such as vegetation monitoring, green infrastructure assessment, and long-term environmental analysis.
\end{itemize}

The rest of this paper is organized as follows. Section~\ref{sec:relatedworks} reviews the related works on the uncertainty issue of multi-modal deep learning tasks and multi-modal semantic segmentation methods. Section~\ref{sec:methodology} describes the MURTreeFormer in detail, including the novel selective uncertainty-guided reconstruction module and the other modules proposed for tree cover feature extraction. Section~\ref{sec:experiments} introduces the datasets, comparative results, and ablation studies. Finally, Section~\ref{sec:conclusion} concludes this paper.

%% file: 2Relatedwork.tex
\subsection{Aleatoric Uncertainty for Multi-modal Fusion}

Aleatoric uncertainty and epistemic uncertainty are two fundamental sources of instability in deep learning models~\cite{kendall2017uncertainties}. While epistemic uncertainty arises from the model’s lack of knowledge and can be reduced with more data or training, aleatoric uncertainty—also known as data uncertainty—originates from inherent noise, ambiguities, or inconsistencies in the input data. In this work, uncertainty caused by multi-modal discrepancies, such as geographic shifts or differences in capture time, falls into the category of the latter.

Effectively modeling aleatoric uncertainty is critical for enhancing both the reliability and interpretability of multi-modal learning systems. Moreover, accurate uncertainty modeling also facilitates robust cross-modal feature fusion. This problem has been explored in various multi-modal tasks, including image-text retrieval, audio-visual speech recognition, and vision-language navigation. Probabilistic models~\cite{abdar2021review}, particularly those employing distributional latent representations~\cite{ji2023map,pmlr-v202-zhang23ar}, have been widely adopted to enhance robustness in multi-modal interactions. For example,~\citet{gao2024embracing} adjusted the relative weights of each modality based on uncertainty estimation to optimize overall decision-making.~\citet{peng2022balanced} adaptively modulated gradient contributions across modalities by monitoring their alignment with the learning objective.%~\citet{wu2022characterizing} introduced conditional utilization rates to quantify how much each modality was used during training and proposed a gradient-derived conditional learning speed to reflect the update dynamics between uni-modal and multi-modal learning.

However, these methods primarily focus on balancing modality contributions or performing decision-level fusion, and are not well suited for dense prediction tasks such as semantic segmentation. More importantly, they fail to address the spatial inconsistencies and data-level heterogeneity between modalities, which are particularly pronounced in high-resolution remote sensing data. In addition, their uncertainty modeling strategies are often designed for classification or retrieval tasks, making them difficult to adapt to pixel-level segmentation problems.

\begin{figure*}[hptb]
  \centering
  \includegraphics[width=16cm]{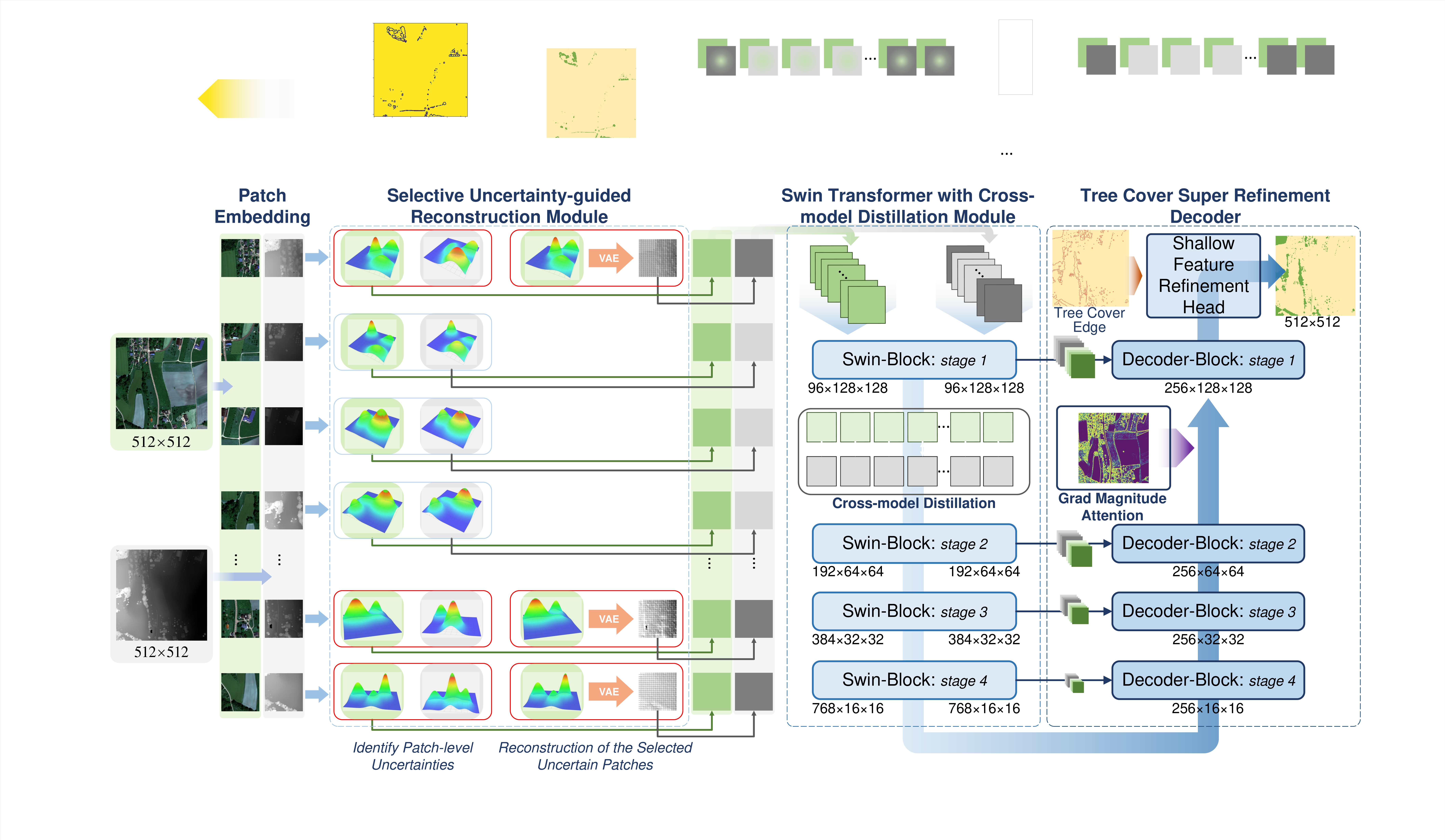}
  \caption{The overall architecture of MURTreeFormer. MURTreeFormer contains a selective uncertainty-guided reconstruction module (SURM), a Swin Transformer based encoder with a cross-model distillation module (CDM), and a tree cover super refinement decoder including grad magnitude attention (GMA) and refinement head (RH).}
  \label{fig:MURTreeFormer}
 \end{figure*}

\subsection{Multi-modal Semantic Segmentation}

Multi-modal semantic segmentation has received increasing attention in recent years for its ability to integrate complementary information from diverse data sources. Such techniques have been widely applied in fields like autonomous driving~\cite{li2022rgb, zhao2023mitigating}, medical image analysis~\cite{yang2022toward, sun2024semi}, and other domains. Deep learning-based fusion encoders have shown great potential in capturing cross-modal correlations, and when combined with suitable decoders, they form powerful end-to-end segmentation frameworks. For instance,~\citet{jia2024geminifusion} proposed GeminiFusion, a linear-complexity pixel-wise fusion network that harmonizes intra- and inter-modal attentions through layer-adaptive noise control, achieving efficient cross-modal integration.

In the field of remote sensing, a number of multi-modal segmentation models have been developed in recent years. TransFusion~\cite{maiti2023transfusion} enabled direct fusion of images and point clouds for semantic segmentation without lossy preprocessing, effectively capturing spatial and geometric information. \citet{zhang2024asanet} introduced ASANet (Asymmetric Semantic Aligning Network), which incorporates feature-level asymmetry to address the common issue that many existing architectures fail to fully exploit complementary cross-modal cues.

However, most of these remote sensing approaches are trained and evaluated on publicly curated datasets, where data quality is typically well controlled, thereby minimizing the impact of uncertainty on segmentation performance. In real-world scenarios, multi-modal data often contain unverified or inconsistent regions due to acquisition limitations or temporal variations, yet this uncertainty is rarely addressed due to annotation or processing costs. Moreover, existing models are not specifically tailored for tree-cover segmentation, making them less effective in capturing the fine-grained structural characteristics of trees in high-resolution remote sensing imagery.

%% file: 3Methodology.tex
The proposed MURTreeFormer is a semantic segmentation network based on deep learning, following a standard encoder-decoder architecture. The encoder is built upon the Swin Transformer~\cite{liu2021swin}, while a progressive decoder is adopted to refine the tree cover segmentation result. The overall architecture of MURTreeFormer is illustrated in Fig.~\ref{fig:MURTreeFormer}. Given a primary modality image $\mathcal{X}_P$ and an auxiliary modality image $\mathcal{X}_A$, MURTreeFormer distinguishes itself from conventional multi-modal segmentation frameworks by incorporating a selective uncertainty-guided reconstruction (SURM) module prior to the encoder. This module is designed to 1) identify patch-level uncertainties across modalities and 2) reconstruct the selected uncertainty patches of the auxiliary modality based on the learned distribution of the primary modality.

\subsection{Selective Uncertainty-guided Reconstruction Module (SURM)}
\label{sec:SURM}
Aleatoric uncertainty is inevitable between cross-temporal multi-modal remote sensing images, compromising high-resolution tree cover semantic segmentation accuracy. To address this, the Selective Uncertainty-guided Reconstruction Module (SURM) is proposed to automatically identify uncertainty patches via a probabilistic framework and rectify them through a variational autoencoder (VAE)~\cite{kingma2014auto} based module. 

During patch embedding, $\mathcal{X}_P$ and $\mathcal{X}_A$ are both divided into $N$ non-overlapping patches and then embedded, where the $i$th embedding of $\mathcal{X}_M$ is denoted as $\mathbf{x}_M^i,\,M\in\left\{P,A\right\}$. SURM creatively quantizes and reconstructs uncertainty at the patch level, which greatly reduces the amount of calculation.

Drawing inspiration from distributional multi-modal feature representation~\cite{gao2024embracing}, mutual-independent multi-variate Gaussian distribution is assigned to the feature $\mathbf{z}_M^i$ conditioned on $\mathbf{x}_M^i$ to achieve fuzzy representation for uncertainty modeling, \textit{i.e.,} 
\begin{equation}
\begin{aligned}
    \mathsf{p}\left(\mathbf{z}_M^i|\mathbf{x}_M^i\right)=\mathcal{N}\left(\mathbf{\mu}_M^i,\mathsf{diag}(\mathbf{\sigma}_M^i)\right),\,M\in\left\{P,A\right\},\\i=1,\cdots,N
\end{aligned}
\end{equation}
where $\mathcal{N}\left(\mathbf{\mu}_M^i,\mathsf{diag}(\mathbf{\sigma}_M^i)\right)$ denotes the multi-variate Gaussian distribution with $\mathbf{\mu}_M^i$ being the mean vector and $\mathsf{diag}(\cdot)$ diagonalizes the variance vector $\mathbf{\sigma}_M^i\succ\mathbf{0}$ to form the co-variance matrix. Thereupon, to characterize $\mathbf{z}_M^i$, it only requires to learn the distributional parameters $\mathbf{\mu}_M^i$ and $\mathbf{\sigma}_M^i$. For this goal, two lightweight shallow extractors~($\mathsf{Le}$), denoted as $\mathsf{Le}_1\left(\cdot;\mathbf{\theta}_1^M\right),\,\mathsf{Le}_2\left(\cdot;\mathbf{\theta}_2^M\right)$ and parameterized with $\mathbf{\theta}_1^M,\,\mathbf{\theta}_2^M$, are employed to deliver the mean and co-variance:
\begin{equation}
	\mathbf{\mu}_M^i=\mathsf{Le}_1\left(\mathbf{x}_M^i;\mathbf{\theta}_1^M\right),\,\ln\left(\mathbf{\sigma}_M^i\right)=\mathsf{Le}_2\left(\mathbf{x}_M^i;\mathbf{\theta}_2^M\right)
\end{equation}
in which the output of $\mathsf{Le}_2\left(\cdot;\mathbf{\theta}_2^M\right)$ is defined as the component-wise logarithm of $\mathbf{\sigma}_M^i$ to ensure the positive definiteness of co-variance matrix.

\begin{table}[htpb]

    \caption{\label{tab:symbol}
    %Some notations.
    Notations and operations used in this study. 
    }
    \setlength{\tabcolsep}{2mm}{
    \begin{tabular}{ll} 
\hline
Notation & Explanation    \\ \hline

$f\!c$         & \begin{tabular}[c]{@{}l@{}}Fully connection operation, which operates independently\\ at each spatial location, and is functionally equivalent to \\a $1 \times 1$ convolution with stride 1.\end{tabular} \\ 

\specialrule{0em}{3pt}{3pt}

$ReLU$ &  \begin{tabular}[c]{@{}l@{}}Rectified linear unit, used as an activation function.\end{tabular}\\ 
    \specialrule{0em}{3pt}{3pt}
    
feature map   &   \begin{tabular}[c]{@{}l@{}}A \emph{width} $\times$ \emph{height} $\times$ \emph{channel} dimensional tensor.\\ Generally, the feature map has three dimensions, with\\ the horizontal and vertical dimensions denoted as\\ width and height, and the last dimension as channel. \\The sizes of \emph{width}, \emph{height} and \emph{channel} are represented\\ by \emph{W}, \emph{H} and \emph{C}, respectively. 
\end{tabular}\\
    \specialrule{0em}{3pt}{3pt}

$conv_{k\times k}$ & \begin{tabular}[c]{@{}l@{}}The convolution operation with a \textit{k} $\times$ \textit{k} sized kernel;\\ the default value of \textit{k} is 3.\end{tabular}\\

\specialrule{0em}{3pt}{3pt}

\textit{BN}  &   \begin{tabular}[c]{@{}l@{}}Batch normalization operation, a basic normalization \\operation.\end{tabular}\\ 
    \specialrule{0em}{3pt}{3pt}

\textit{sigmoid}  &  \begin{tabular}[c]{@{}l@{}}Sigmoid function with the equation $\frac{1}{{1 + {e^{ - x}}}}$, used as an \\activation function.\end{tabular}\\ 
\specialrule{0em}{3pt}{3pt}

\textit{softmax}  &  \begin{tabular}[c]{@{}l@{}}Softmax function with the equation $\frac{e^{x_i}}{\sum_{j} e^{x_j}}$, used to\\ convert logits into a probability distribution.\end{tabular}\\ 
\specialrule{0em}{3pt}{3pt}

\textit{gp}     &  \begin{tabular}[c]{@{}l@{}}Global pooling operation, which selects the maximum\\ value from each \emph{W}$\times$\emph{H} region; the size of the output\\ feature map is 1 $\times$ 1 $\times$ \emph{C}.\\
\end{tabular}\\ 
\specialrule{0em}{3pt}{3pt}

\textit{concate}         & \begin{tabular}[c]{@{}l@{}}Concatenate operation. For the two feature maps\\ $f_1\in \mathbb{R} ^{W \times H \times C_1}$ and $f_2\in \mathbb{R} ^{W \times H \times C_2}$, the operation\\ yields $[f_1,f_2] \in \mathbb{R} ^{W \times H \times (C_1+C_2)}$\end{tabular} \\ 

\hline
\end{tabular}}
\end{table}

\subsubsection{Identification of the Patch-level Uncertainties}

To identify spatially uncertain regions for patch-wise reconstruction, a learnable patch-wise scoring module (PatchScore) is introduced to evaluate the inconsistency between modalities. Given the patch embeddings $\mathbf{x}_P^i$ and $\mathbf{x}_A^i$ from the primary and auxiliary modalities, along with their latent distributional representations $\mathsf{p}(\mathbf{z}_P^i|\mathbf{x}_P^i)$ and $\mathsf{p}(\mathbf{z}_A^i|\mathbf{x}_A^i)$, the patch-level inconsistency score is computed as follows:

\begin{equation}
\begin{aligned}
    v^i &= \frac{1}{2}\ln\left(\frac{\left|\mathsf{diag}(\mathbf{\sigma}_P^i)\right|}{\left|\mathsf{diag}(\mathbf{\sigma}_A^i)\right|}\right),\,i=1,\cdots,N\\
    RawM\!ap &= f\!c(ReLU(f\!c([v^1,v^2,\cdots,v^N]))), \\
    ScoreM\!ap &= so\! ftmax(RawM\!ap), \\
\end{aligned}
\end{equation}
where $\left|\cdot\right|$ stands for the determinant absolute value of a square matrix. $ v^{i}$ is then the entropy difference between the primary and auxiliary latent features given the entropy of Gaussian distributions \cite{MISRA2005324},
\iffalse
\begin{align}
&-\mathsf{E}\left\{\ln\left(\mathsf{p}\left(\mathbf{z}_M^i|\mathbf{x}_M^i\right)\right)\right\}\notag\\
=&-\mathsf{E}\left\{\ln\left(\frac{e^{-(\mathbf{z}_M^i-\mathbf{\mu}_M^i)^T(\mathsf{diag}(\mathbf{\sigma}_M^i))^{-1}(\mathbf{z}_M^i-\mathbf{\mu}_M^i)/2}}{(2\pi)^{d/2}\left|\mathsf{diag}(\mathbf{\sigma}_M^i)\right|^{1/2}}\right)\right\}\notag\\=&-\mathsf{E}\Big\{-\frac{1}{2}\Big(d\ln(2\pi)+\ln(\left|\mathsf{diag}(\mathbf{\sigma}_M^i)\right|\notag\\&+(\mathbf{z}_M^i-\mathbf{\mu}_M^i)^T(\mathsf{diag}(\mathbf{\sigma}_M^i))^{-1}(\mathbf{z}_M^i-\mathbf{\mu}_M^i)\Big)\Big\}\notag\\=&\frac{1}{2}\left(d\ln(2\pi)+\ln(\left|\mathsf{diag}(\mathbf{\sigma}_M^i)\right|+d\right)\notag\\=&\frac{1}{2}\ln((2\pi e)^d\left|\mathsf{diag}(\mathbf{\sigma}_M^i)\right|)
\end{align}
where $\mathbf{z}_M^i$ is assumed $d$-dimensional and $\mathsf{E}\left\{\cdot\right\}$ denotes the expectation operator, thereupon
\begin{align}
v^i&=\frac{1}{2}\ln((2\pi e)^d\left|\mathsf{diag}(\mathbf{\sigma}_P^i)\right|)-\frac{1}{2}\ln((2\pi e)^d\left|\mathsf{diag}(\mathbf{\sigma}_A^i)\right|)\notag\\&=-\mathsf{E}\left\{\ln\left(\mathsf{p}\left(\mathbf{z}_P^i|\mathbf{x}_P^i\right)\right)\right\}-\left(-\mathsf{E}\left\{\ln\left(\mathsf{p}\left(\mathbf{z}_A^i|\mathbf{x}_A^i\right)\right)\right\}\right).
\end{align}
\fi
which serves as an initial proxy for distributional inconsistency. A lightweight two-layer MLP~\cite{tolstikhin2021mlp} is then applied to refine these raw scores, and a softmax operation is used to normalize the output into the final $ ScoreM\!ap\triangleq\left[s^1,\cdots,s^N\right]$ where $s^i$ represents the uncertainty score of the $i$th patch.

$ScoreM\!ap$ reflects the semantic inconsistency between corresponding patches across modalities. For each input group, SURM obtains a $K$-element set $\mathbb{T}\subset\left\{1,\cdots,N\right\}$ indexing the top-$K$ patches with the highest inconsistency uncertainty scores, \textit{i.e.}, the largest $K$ score values of $s^{i},\,i=1,2,\cdots,N$.

In practice, for a $512 \times 512$ input image yielding $N=16384$ patches with size $4 \times 4$, only $K=500$ patches are selected for reconstruction. This significantly reduces the computational overhead compared to full-map refinement and focuses the reconstruction on cross-modal inconsistent uncertainty regions. The final uncertainty map $ScoreM\!ap$ can also be visualized to interpret the uncertainty distribution spatially.

\subsubsection{Reconstruction of the Selected Uncertain Patches}
\label{sec:reconstruction}
\begin{figure}[htbp]
  \centering
  \includegraphics[width=8cm]{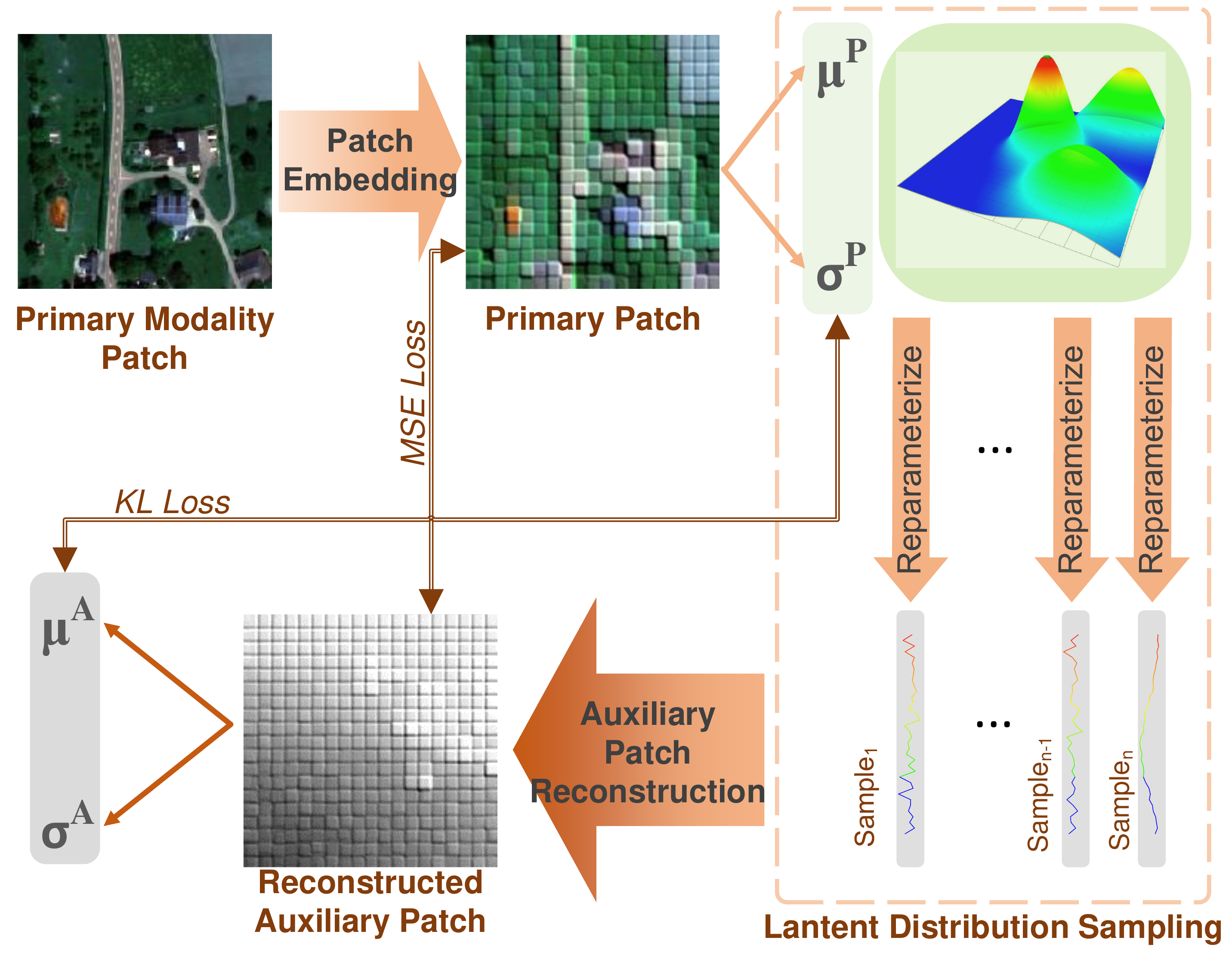}
  \caption{Illustration of the patch-wise reconstruction module (PatchRecon).}
  \label{fig:PatchReconstructModule}
 \end{figure}

To ensure a probabilistically grounded and uncertainty-aware reconstruction process, MURTreeFormer incorporates a VAE-based component named the patch-wise reconstruction module (PatchRecon). In this design, the latent distributions of the primary modality serve as priors to generate robust auxiliary features via reparameterization and decoding. This promotes the learning of smooth and expressive latent manifolds, which are particularly beneficial in ambiguous regions.

Traditional VAE-based reconstruction methods often produce blurry outputs and lack spatial precision when applied to dense prediction tasks like semantic segmentation. This is mainly because global reconstruction dilutes the supervision across the entire image, making it harder to learn fine-grained structures such as object boundaries and textures. In addition, image-level uncertainty modeling lacks spatial specificity, which limits its ability to target semantically ambiguous areas. To overcome these challenges, PatchRecon performs localized reconstruction by selectively focusing on high-uncertainty patches. This strategy concentrates learning on structurally complex and informative regions, resulting in more accurate and robust multi-modal feature representations.

Specifically, after the $K$-element set $\mathbb{T}\subset\left\{1,\cdots,N\right\}$ with the highest inconsistency uncertainty scores being selected, for each patch index $i \in \mathbb{T}$, $L$ latent samples $\widetilde{\mathbf{z}}_P^{i,j},\,j=1,\cdots,L,$ are generated from the learned distributional parameters of the primary modality via the reparameterization trick:
\begin{equation}
    \widetilde{\mathbf{z}}_P^{i,j} = \mathbf{\mu}_P^i + \mathbf{\sigma}_P^i \odot \mathbf{\epsilon}^j, \quad \mathbf{\epsilon}^j \sim \mathcal{N}(\mathbf{0}, \mathbf{I}),
\end{equation}
where $\odot$ denotes the Hadamard product and $\mathbf{\epsilon}^j$ is randomly sampled from the standard Gaussian distribution. 
A lightweight decoder is then used to obtain the final reconstructed auxiliary patch:
\begin{equation}
\begin{aligned}
    \hat{\mathbf{x}}_A^i = f\!c(ReLU(f\!c([ \widetilde{\mathbf{z}}_P^{i,1},\cdots, \widetilde{\mathbf{z}}_P^{i,L}]))), \quad i \in \mathbb{T}. \\
\end{aligned}
\end{equation}
These reconstructed auxiliary patches are used to replace the original $\mathbf{x}_A^i$ at the selected indices $i \in \mathbb{T}$. 

A mean square error (MSE) loss is applied between the reconstructed $\hat{\mathbf{x}}_A^i$ and the corresponding primary patches $\mathbf{x}_P^i$:
\begin{equation}
    \mathcal{L}_{\text{MSE}} = \frac{1}{K} \sum_{i \in \mathbb{T}} \left\| \hat{\mathbf{x}}_A^i - \mathbf{x}_P^i \right\|^2.
\end{equation}
In addition, a Kullback-Leibler (KL) divergence loss is applied between the distributional parameters of the auxiliary and primary modalities:
\begin{equation}
    \mathcal{L}_{\text{KL}} = \sum_{i\in\mathbb{T}}D_{\text{KL}}\left[ \mathcal{N}(\mathbf{\mu}_A^i, \mathsf{diag}(\mathbf{\sigma}_A^i)) \,\|\, \mathcal{N}(\mathbf{\mu}_P^i, \mathsf{diag}(\mathbf{\sigma}_P^i)) \right].
\end{equation}

For spatial coherence maintenance and downstream fusion compatibility, the original auxiliary patch feature $\mathbf{x}_A^i$ is selectively replaced by its reconstructed version according to the index set $\mathbb{T}$:
\begin{equation}
    \mathbf{x}_A^i \leftarrow 
    \begin{cases}
        \hat{\mathbf{x}}_A^i, & \text{if } i \in \mathbb{T}, \\
        \mathbf{x}_A^i, & \text{otherwise}.
    \end{cases}
\end{equation}
Through this conditional replacement strategy, native features of the auxiliary modality are preserved in reliable regions, while feature consistency in uncertain areas is enhanced via cross-modal reconstruction. The refined auxiliary patch features $\left\{ \mathbf{x}_A^i \right\}_{i=1}^N$ are subsequently reassembled into the complete auxiliary feature map and propagated to the next steps.

\subsection{Cross-modal Distillation Module}
\label{sec:CDM}
To alleviate global discrepancies between heterogeneous modalities, the cross-modal distillation module (CDM) is introduced. It aligns the auxiliary modality’s latent features to the primary modality using a distribution-level loss. Inspired by the concept of knowledge distillation, CDM enables the proposed model to learn more consistent and reliable representations from the auxiliary modality under the guidance of the primary one. CDM is intentionally positioned at stage~2 of Swin Transformer based encoder as illustrated in Fig.~\ref{fig:MURTreeFormer}. This choice is motivated by the fact that very shallow features are dominated by modality-specific textures and noise, which may hinder effective alignment. At stage~1, the features already incorporate preliminary semantic context while still retaining spatial granularity, making them more suitable for cross-modal alignment.

Following the SURM process, each modality is represented as a sequence of $N$ patch embeddings, matrix-wisely denoted as $\mathbf{X}_P\triangleq\left[\mathbf{x}_P^1,\cdots,\mathbf{x}_P^N\right], \mathbf{X}_A\triangleq\left[\mathbf{x}_A^1,\cdots,\mathbf{x}_A^N\right]$. After feature extraction via Swin Transformer blocks (Swin-Block), each $\mathbf{x}_M^i$, where $M \in \{P, A\}$, represents a patch-level feature vector enriched with local spatial and semantic information. 

To align these modality-specific features in a shared representation space, the patch sequences are first passed through modality-specific embedding heads~\cite{tsai2019multimodal}, implemented as two parallel learnable fully connected layers:
\[
\tilde{\mathbf{X}}_P = \mathsf{Proj}_P(\mathbf{X}_P), \quad 
\tilde{\mathbf{X}}_A = \mathsf{Proj}_A(\mathbf{X}_A),
\]
where $\mathsf{Proj}_P$ and $\mathsf{Proj}_A$ independently project the primary and auxiliary modality features into a unified latent space for cross-modal alignment.

To encourage directional consistency between the two modalities, the cosine similarity is computed between the projected embeddings, and the following distillation loss is minimized:
\begin{equation}
\mathcal{L}_{\mathrm{CDM}} = 1 - \cos(\tilde{\mathbf{X}}_P, \tilde{\mathbf{X}}_A).
\end{equation}
Cosine distance is chosen for its scale-invariant property and for providing stable gradients during patch-level supervision. This enables smoother training and facilitates better alignment of high-dimensional token directions, especially in structurally complex regions. 

During training, $\mathcal{L}_{\mathrm{CDM}}$ is optimized jointly with the segmentation objective, allowing the encoder to gradually regularize the features from both modalities toward a modality-invariant embedding space. As a result, CDM serves as a global alignment constraint that enhances cross-modal consistency and benefits the downstream decoding process.

\subsection{Tree Cover Super Refinement Decoder}

\begin{figure}[htbp]
  \centering
  \includegraphics[width=9cm]{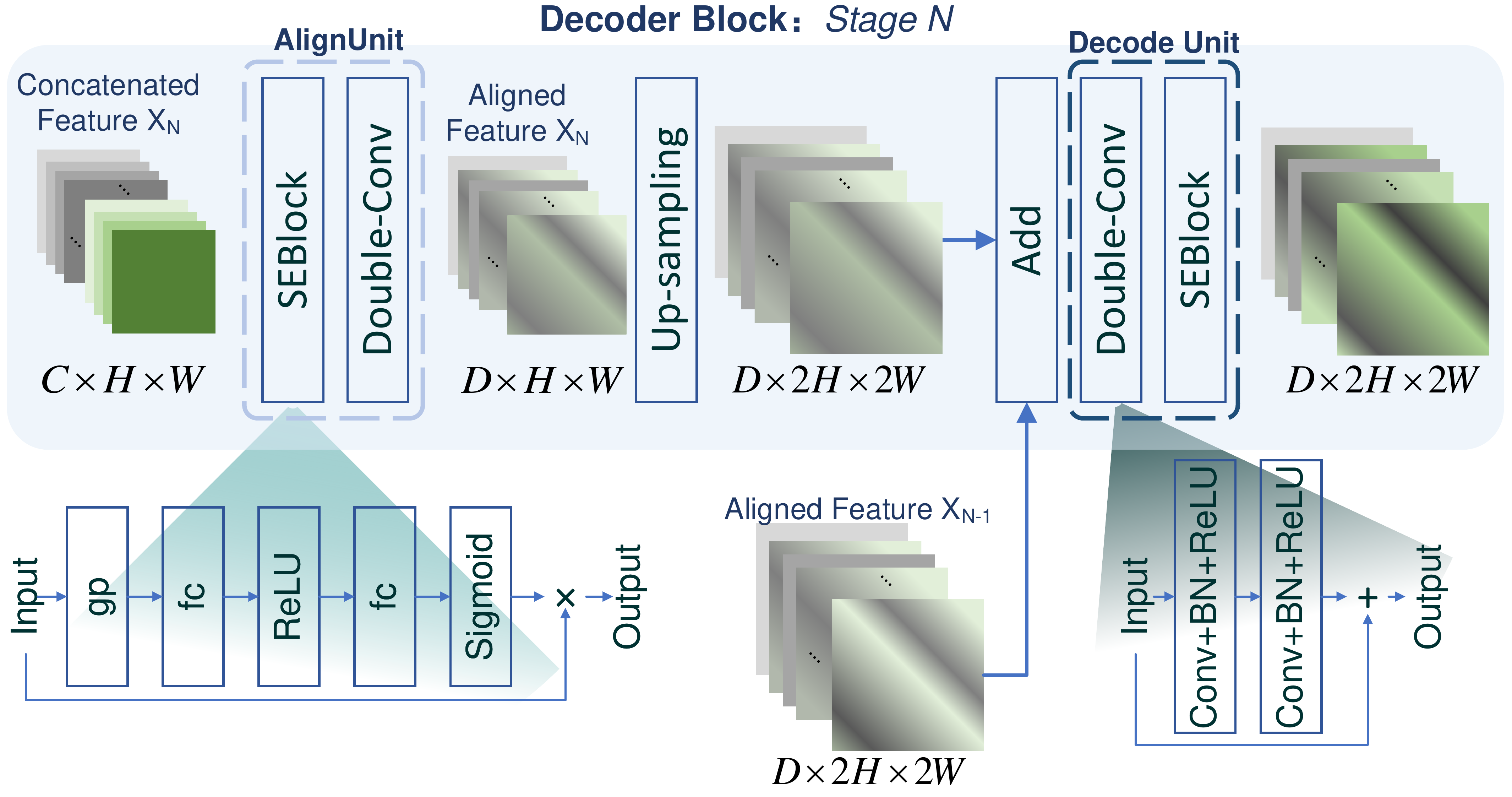}
  \caption{Illustration of the decoder block. Each stage consists of an AlignUnit and a DecoderUnit. The AlignUnit is responsible for reducing the channel dimension and aligning cross-modal features, while the DecoderUnit is applied to progressively restore fine-grained segmentation details.}
  \label{fig:DecoderBlock}
 \end{figure}
 
To accurately decode tree cover regions with varying spatial scales, a tree cover super refinement decoder is proposed to progressively integrate hierarchical features while enhancing spatial awareness and refining spatial resolution. Based on the progressive decoder, two modules are designed for the tree cover class: 1) a gradient magnitude attention (GMA) for enhancing visually coherent structures, and 2) a shallow feature refinement head (RH) that restores hierarchical semantic details from coarse to fine.

The basic decoder block is shown in Fig.~\ref{fig:DecoderBlock}, which consists of an AlignUnit and a DecoderUnit, operating at different resolution stages. Since the feature map outputs from different Swin-Block stages have different channel dimensions and originate from different modalities, normalization and alignment are necessary before decoding. As such, the AlignUnit is first applied to compress the channel dimensions and harmonize cross-modal features. Specifically, modality-specific features are concatenated and passed through an attention-guided block that incorporates SE Block~\cite{hu2018squeeze}, enabling channel-wise recalibration based on global context. This step ensures that heterogeneous features are adaptively fused and aligned for subsequent decoding. Next, the DecoderUnit processes the aligned features by gradually upsampling and refining them. Each decoder unit contains a double convolution block followed by an SE Block, enabling it to capture spatial structure while selectively emphasizing informative features. This progressive design allows tree-like objects and boundary information to be recovered more accurately.

The decoder block functions as a stage-wise refinement pipeline that fuses, aligns, and decodes multi-modal features in a resolution-aware manner, effectively enhancing segmentation robustness and detail sensitivity. In addition, two other modules are designed to further refine the segmentation results of high-resolution tree cover:

\begin{figure}[htbp]
  \centering
  \includegraphics[width=8cm]{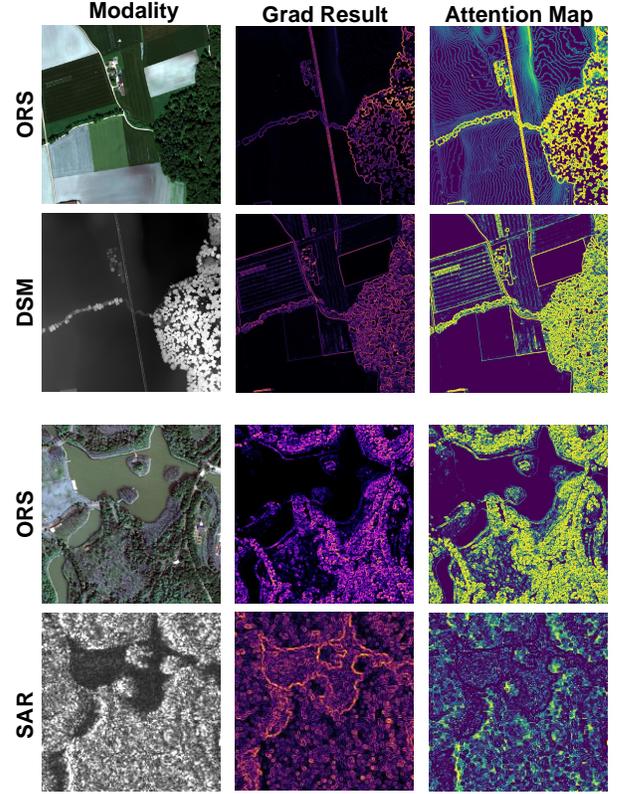}
  \caption{Visual illustration of gradient-based attention maps generated by the proposed gradient magnitude attention (GMA). Two sets of input examples are shown, covering ORS, DSM, and SAR modalities. 
  GMA can capture tree cover boundaries with high spatial precision across modalities, producing visually consistent attention responses despite large spectral and structural differences.
  }
  \label{fig:GAM_effect}
 \end{figure}

\subsubsection{Gradient Magnitude Attention}

Accurately segmenting object boundaries remains a persistent challenge in semantic segmentation, especially for tree cover in remote sensing imagery, which typically exhibits irregular and fragmented edges. To address this, we introduce the gradient magnitude attention (GMA) module, an effective attention mechanism designed to enhance feature perception along luminance-continuous regions. GMA is compatible with multiple modalities and guides the decoder's focus toward structurally informative areas, such as tree boundaries that often present subtle visual cues. 

Given a primary modality image $\mathcal{M}_P$ (e.g., an optical image with red, green, and blue spectral bands), a luminance map $\mathbf{L}_{{opt}}$ is first computed by averaging the RGB channels:

\begin{equation}
    \mathbf{L}_{opt} = \frac{1}{3}({Red} + {Green} + {Blue}).
\end{equation}
To amplify contrast in dark forested regions, the luminance map is inverted, normalized, and then enhanced via luminance attention module (LAM)~\cite{gui2025mapping}:
\begin{equation}
    \bar{\mathbf{L}} = 1 - \frac{\mathbf{L}_{opt}}{\max(\mathbf{L}_{opt})}, \quad
    \tilde{\mathbf{L}} = \bar{\mathbf{L}}^\gamma, \quad \gamma > 1,
\end{equation}
where the power operation $\gamma$ is component-wise.
Next, Sobel filters are used to compute the luminance gradient, capturing edge intensity in both horizontal and vertical directions:
\begin{equation}
\nabla{\mathbf{L}} = \sqrt{(G_x \ast \tilde{\mathbf{L}})^2 + (G_y \ast \tilde{\mathbf{L}})^2},
\end{equation}
where $G_x$ and $G_y$ are Sobel kernels, and $\ast$ denotes convolution. The resulting gradient map $\nabla{\mathbf{L}}\in[0,1]$, where larger values correspond to sharper edge transitions. To emphasize these high-gradient areas (such as tree boundaries), an exponential decay function is applied to suppress large values and compress the dynamic range. The result is then inverted by subtracting from 1, producing an attention map where regions with stronger edges receive higher weights:
\begin{equation}
    \mathbf{A} = 1- \exp(-\nabla{\mathbf{L}}).
    \label{eq:gradatten}
\end{equation}

Considering the uncertainty of multi-modal inputs, MURTreeFormer performs reconstruction based on the primary modality. Accordingly, the attention map in GMA is derived from the primary modality to more effectively guide the refinement process in the decoder.

As illustrated in Fig.~\ref{fig:GAM_effect}, the attention maps extracted using GMA from different modalities consistently highlight tree cover boundaries. For ORS and DSM modalities, GMA produces sharp and well-localized responses along vegetation edges. Although attention maps derived from SAR images are more spatially dispersed, GMA still extracts informative texture cues. Visual inspection suggests that in SAR, the attention maps may be more effectively interpreted as suppression priors, helping the decoder ignore noisy or irrelevant regions.

\subsubsection{Shallow Feature Refinement Head}

After the four-stage decoder, the feature maps are upsampled to a spatial resolution of $\frac{H}{4} \times \frac{W}{4}$. Instead of applying a single-step upsampling as in conventional Transformer-based decoders~\cite{xie2021segformer}, we introduce a dedicated shallow feature refinement head (RH) to progressively recover fine-grained spatial details.
Unlike the commonly adopted auxiliary heads~\cite{zhao2017pyramid}, which are used solely during training for intermediate supervision, the proposed RH is a lightweight yet effective decoder component that remains active during inference. It explicitly enhances feature granularity through progressive upsampling and multi-task supervision. By jointly predicting both segmentation and edge maps, RH improves boundary precision and reinforces spatial consistency—an essential requirement for accurate tree cover delineation in high-resolution remote sensing imagery.

\begin{figure}[htbp]
  \centering
  \includegraphics[width=7cm]{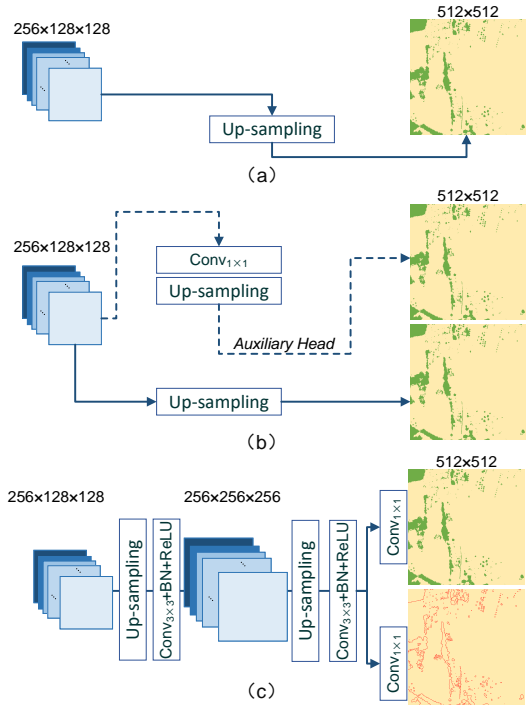}
  \caption{Comparison of decoder heads. 
(a) Single-step upsampling head commonly used in Transformer-based segmentation models such as SegFormer. 
(b) Auxiliary head, typically used during training for deep supervision. 
(c) The proposed shallow feature refinement head (RH), which performs progressive upsampling and jointly predicts segmentation and edge maps to enhance spatial consistency.
  }
  \label{fig:decoder_comparision}
 \end{figure}

Abrupt upsampling may lead to blurred boundaries and loss of subtle structures, especially in remote sensing scenes where tree cover boundaries often exhibit fragmented textures and irregular contours. Instead of a one-shot interpolation, RH adopts a two-step progressive upsampling strategy to gradually refine spatial resolution and preserve geometric continuity.

Specifically, as illustrated in Fig.~\ref{fig:decoder_comparision}~(c), the upsampling path consists of two stages, each comprising bilinear interpolation followed by a $3 \times 3$ convolution, batch normalization, and ReLU activation. To enhance supervision and guide the network toward preserving fine structural details, two prediction heads are introduced: one generates the final semantic segmentation output, while the other produces an auxiliary prediction supervised by edge annotations. By explicitly modeling edge information, RH helps the decoder better delineate fragmented or narrow tree crowns. In addition, this design mitigates feature degradation caused by repeated upsampling operations and promotes improved spatial consistency.

%% file: 4Experiments.tex
The study areas for this research are Zurich (Switzerland) and Shanghai (China). Comprehensive experiments are conducted on two multi-modal datasets to evaluate the tree cover segmentation performance of the proposed MURTreeFormer. In this section, the two datasets are described in detail, followed by the experimental settings, loss functions, and evaluation metrics. The performance of MURTreeFormer is then compared with several state-of-the-art methods, and detailed ablation studies are presented to validate the effectiveness of each component.

\begin{table}[htbp]
\caption{Dataset Information for Tree Cover Segmentation}
\label{tab:dataset_detail}
\resizebox{8cm}{!}{ % Adjust the width to 8.5cm, keeping the height aspect ratio
\begin{tabular}{c|ccccc}
\hline
\multirow{2}{*}{Dataset}  & \multirow{2}{*}{Modality} & \multirow{2}{*}{Source}       & \multirow{2}{*}{\begin{tabular}[c]{@{}c@{}}Capture \\ Time\end{tabular}} & \multirow{2}{*}{Resolution} & \multirow{2}{*}{\begin{tabular}[c]{@{}c@{}}Number \\ of Bands\end{tabular}} \\
            & &   &    & &       \\ \hline
\multirow{2}{*}{Shanghai} & ORS         & \begin{tabular}[c]{@{}c@{}}GF-7\\ Satellite\end{tabular}  & 2019-2021        & 0.68m         & 4     \\ \cline{2-6} 
            & SAR         & \begin{tabular}[c]{@{}c@{}}GF-3 \\ Satellite\end{tabular} & 2019-2021        & 3m            & 1     \\ \hline
\multirow{2}{*}{Zurich}   & ORS         & \begin{tabular}[c]{@{}c@{}}Aerial\\  Image\end{tabular}   & 2022 & 0.1m          & 4     \\ \cline{2-6} 
            & DSM         & ALS & 2018 & 1m            & 1     \\ \hline
\end{tabular}

}
\end{table}

\subsection{Experiment Datasets}
\subsubsection{Zurich Dataset}

\begin{figure}[htbp]
  \centering
  \includegraphics[width=8cm]{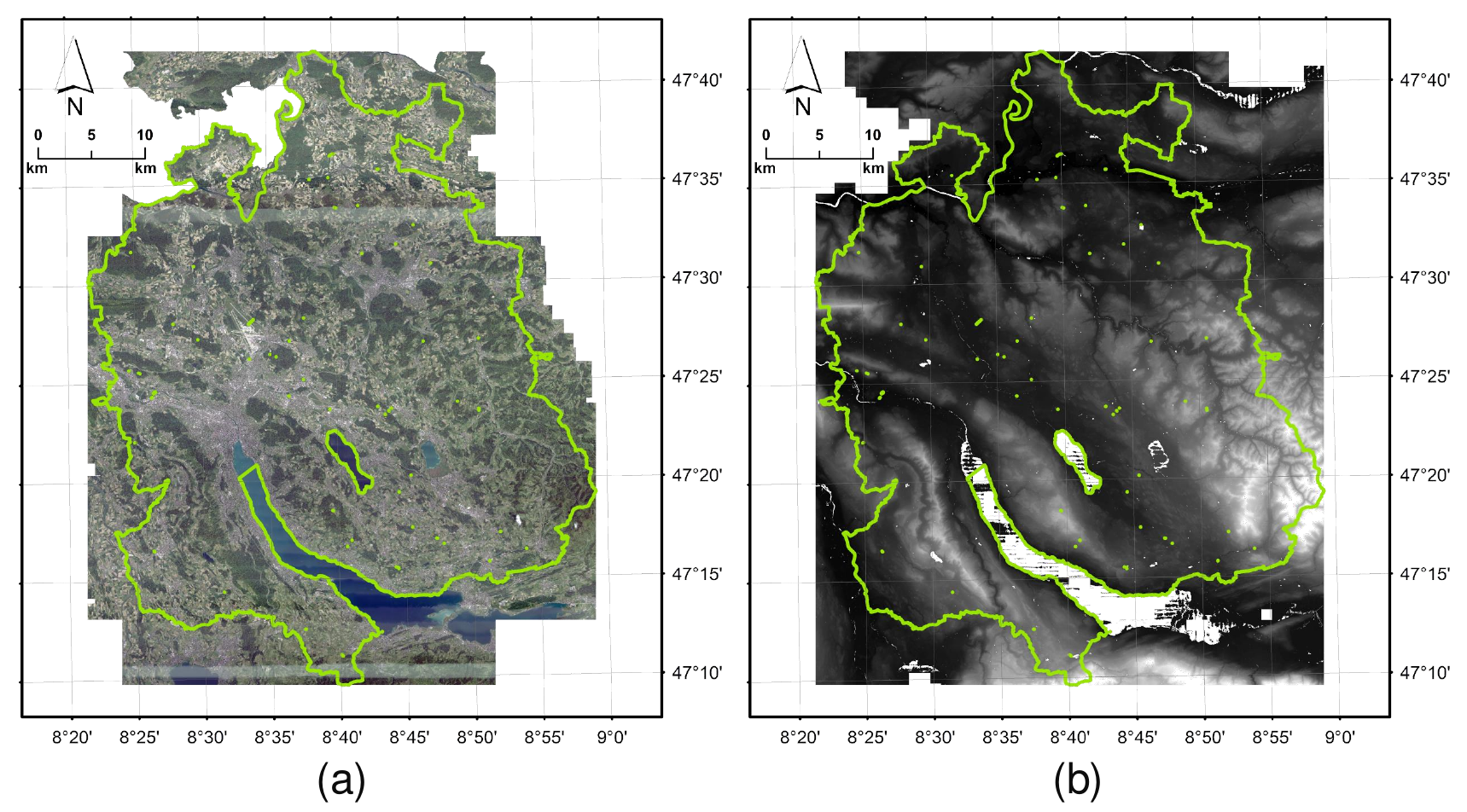}
  \caption{The study area of the Zurich dataset. (a) Optical remote sensing
(ORS) imagery coverage. (b) Digital surface model (DSM) imagery coverage. The green contour indicates the boundary of the study area.
  \label{fig:zurich_dataset}
  }
 \end{figure}
The Zurich dataset covers the canton of Zurich, Switzerland, spanning an area of approximately 1,728.95~km$^2$. The optical remote sensing (ORS) imagery was acquired in 2022 via aerial photography at a spatial resolution of 10~cm, consisting of four spectral bands: red, green, blue, and near-infrared (NIR). Tree cover annotations were automatically generated using the vegetation height model (VHM)~\cite{ginzler2015countrywide}, with a resolution of 1~m. In addition, a 1~m resolution Digital Surface Model (DSM) produced by airborne laser scanning (ALS) in 2018 is included as the auxiliary modality.

To ensure consistency with the label and DSM resolutions, the ORS images are downsampled to 1~m. After geospatial alignment and cropping based on geographic metadata, a total of 9,032 image samples are generated. Each sample comprises a 512$\times$512 patch containing the downsampled ORS image, DSM, and corresponding label. The dataset is divided into training, validation, and test subsets with a ratio of 7:1.5:1.5.

\subsubsection{Shanghai Dataset}

\begin{figure}[htbp]
  \centering
  \includegraphics[width=8cm]{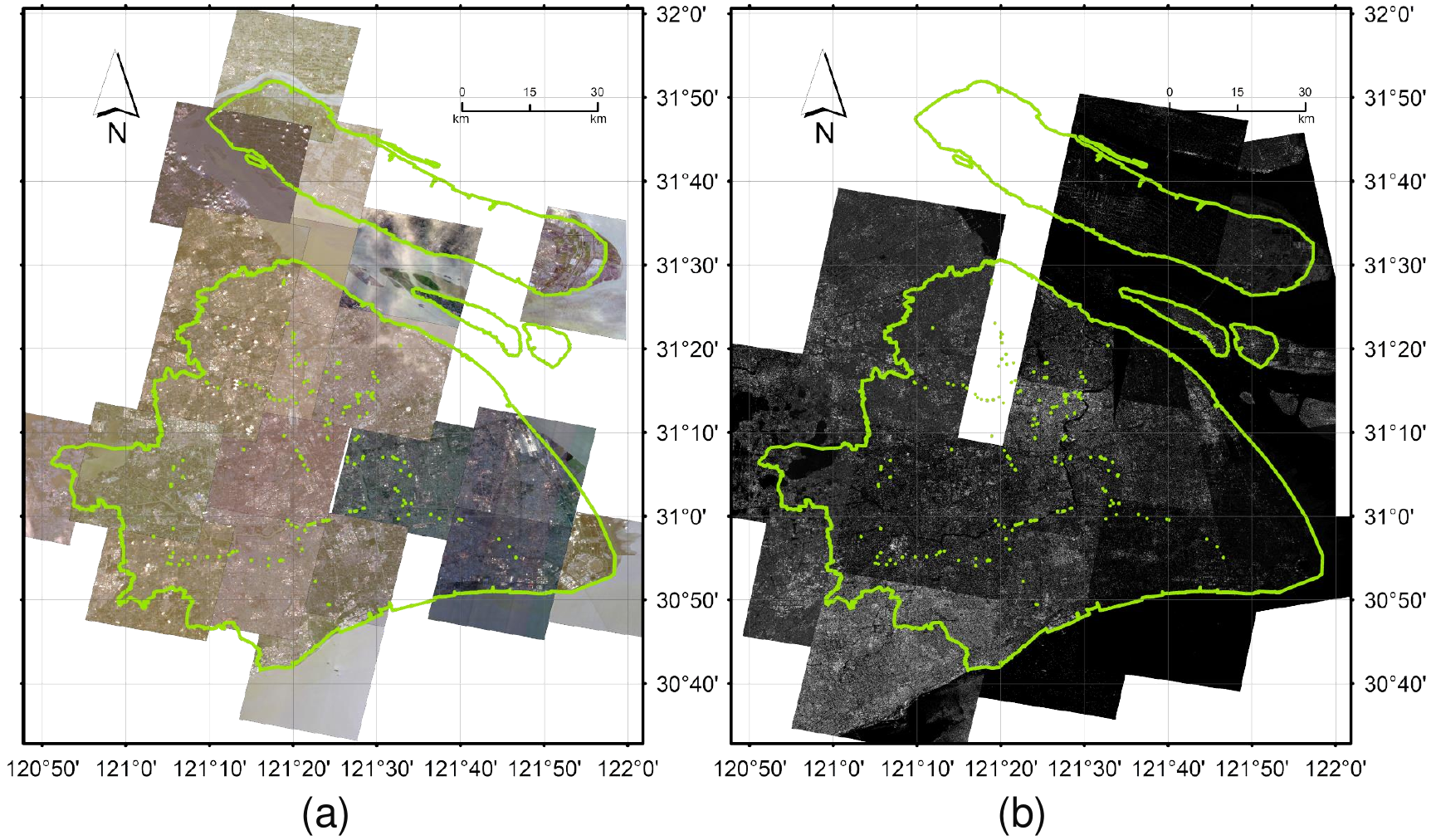}
  \caption{The study area of the Shanghai dataset. (a) Optical remote sensing
(ORS) imagery coverage. (b) Synthetic aperture radar (SAR) imagery coverage. The green contour indicates the boundary of the study area.
  \label{fig:shanghai_dataset}
  }
 \end{figure}

Shanghai dataset encompasses an urban extent of approximately 6,340.5~km$^2$. ORS images from the GF-7 satellite and synthetic aperture radar (SAR) images from the GF-3 satellite were acquired between 2019 and 2022. The pixel-wise tree cover annotations were provided by GeoWit Co., Ltd., based on the GF-7 optical imagery.

The ORS images contain four spectral bands, including red, green, blue, and near-infrared (NIR). The extracted overlapping regions between the optical and SAR images are co-registered and spatially aligned to construct paired multimodal samples. In total, 74,780 image pairs are obtained, each comprising a 0.68~m resolution optical image, a 3~m resolution SAR image, and a 0.68~m resolution ground-truth label. To ensure spatial consistency, SAR images are upsampled to 512$\times$512 pixels via bilinear interpolation to match the resolution and size of the optical and label images.

The dataset is partitioned into training, validation, and test sets in the ratio of 7:1.5:1.5. Compared with the Zurich dataset, the Shanghai dataset has more pronounced temporal gaps and spectral discrepancies (as illustrated in Fig.~\ref{fig:shanghai_dataset}), leading to greater challenges for accurate tree cover segmentation.

\subsubsection{The Cross-modal Uncertainty in Datasets}
Table~\ref{tab:dataset_detail} summarizes the image characteristics of the Zurich and Shanghai datasets. These datasets were obtained at different times using sensors with various modalities, resulting in inherent cross-modal inconsistencies. Such uncertainty becomes more pronounced in high-resolution imagery, where even subtle differences may reduce semantic segmentation performance.

\begin{figure}[htbp]
  \centering
  \includegraphics[width=6cm]{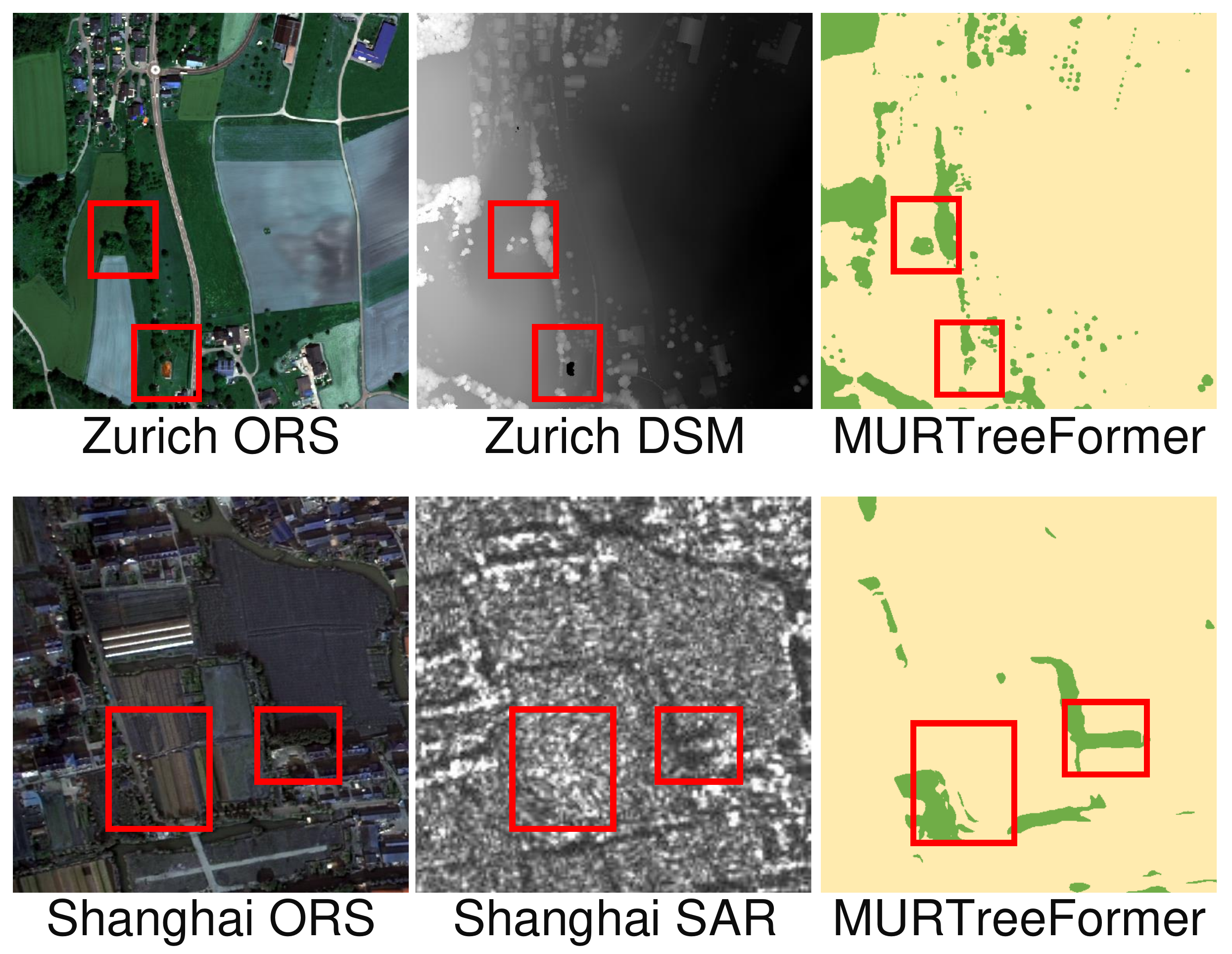}
  \caption{
Illustration of cross-modal uncertainty in the Zurich and Shanghai datasets. The segmentation results from MURTreeFormer demonstrate its ability to leverage reliable information from the primary modality to mitigate the impact of such uncertainty.
}
  \label{fig:uncertainty_example}
 \end{figure}

As shown in Fig.~\ref{fig:uncertainty_example}, the Zurich dataset includes images acquired in different years, during which vegetation changes (e.g., pruning or phenological shifts) lead to mismatches between spectral and elevation information. In the Shanghai dataset, SAR and ORS images do not correspond to each other consistently in vegetated and cultivated areas due to the differences in sensing geometry and acquisition time. These challenges highlight the difficulty of cross-modal uncertainty, which often cannot be solved manually.

\subsection{Loss Function and Setting of Experiments}

The proposed MURTreeFormer produces two outputs: a binary segmentation map for tree cover and a corresponding edge map. To optimize both the main and auxiliary tasks, a multi-objective loss function is adopted. The total training loss is defined as:

\begin{equation}
\mathcal{L}_{\text{total}} = \lambda_{\text{S}} \mathcal{L}_{\text{seg}} + \lambda_{\text{E}} \mathcal{L}_{\text{edge}} + \lambda_{\text{R}} \mathcal{L}_{\text{rec}}  + \lambda_{\text{C}} \mathcal{L}_{\text{CDM}},
\end{equation}
the segmentation loss $\mathcal{L}_{\text{seg}}$ is based on the Intersection over Union (IoU)~\cite{rahman2016optimizing}, which measures the overlap between the predicted region and the label. This loss function directly quantifies the accuracy of the tree cover region delineation. Specifically, it is computed as:

\begin{equation}
\mathcal{L}_{\text{seg}} = 1 - \frac{\text{TP}_{seg}}{\text{TP}_{seg}+\text{FP}_{seg}+\text{FN}_{seg}},
\end{equation}
where $\text{TP}_{seg}$ is the number of the true positive pixels, $\text{FP}_{seg}$ is the number of the false positive pixels and $\text{FN}_{seg}$ is the number of the false positive pixels. For this task, the positive class is tree cover, and the negative class is non-tree.

The edge loss $\mathcal{L}_{\text{edge}}$ is also based on the IoU, but it focuses specifically on optimizing the accuracy of object boundaries. While $\mathcal{L}_{\text{seg}}$ ensures that the overall area is segmented correctly,  $\mathcal{L}_{\text{edge}}$ focuses on fine-tuning the model to recover boundaries, particularly when the object's edges are less distinct or blurred. This is important for delineating fine structures, such as tree crowns and urban boundaries.  $\mathcal{L}_{\text{edge}}$ is computed in a similar formula to $\mathcal{L}_{\text{seg}}$, but it specifically measures the boundaries of objects:
 \begin{equation}
\mathcal{L}_{\text{edge}} = 1 - \frac{\text{TP}_{edge}}{\text{TP}_{edge}+\text{FP}_{edge}+\text{FN}_{edge}},
\end{equation}
where $\text{TP}_{edge}$ is the number of the true positive pixels, $\text{FP}_{edge}$ is the number of the false positive pixels and $\text{FN}_{edge}$ is the number of the false positive pixels. In this task, edge pixels along the boundaries of tree cover regions are considered as positive samples, whereas all other pixels are treated as negative samples.

As described in Section~\ref{sec:SURM}, the reconstruction loss in SURM module consists of two terms: an MSE loss and a KL divergence regularization term derived from the VAE framework. The MSE loss $\mathcal{L}_{\text{MSE}}$ measures the distance between the reconstructed auxiliary modality patches and the original patches within the selected uncertain regions. The KL divergence term $\mathcal{L}_{\text{KL}}$ is defined as the divergence between the approximate posterior distribution and the standard Gaussian prior over the selected latent variables. By minimizing the KL divergence, the encoder is encouraged to produce smooth and expressive latent manifolds, which are critical for stable reparameterization and effective patch-level reconstruction under uncertainty. The overall reconstruction loss is formulated as:
\begin{equation}
\mathcal{L}_{\text{rec}} = \lambda_{\text{mse}}\, \mathcal{L}_{\text{MSE}} + \lambda_{\text{kl}}\, \mathcal{L}_{\text{KL}},
\end{equation}
where $\lambda_{\text{mse}}$ and $\lambda_{\text{kl}}$ are weighting coefficients balancing the contribution of each term.

The cross-modal distillation loss $\mathcal{L}_{\text{CDM}}$ is designed to encourage representation consistency between the primary and auxiliary modalities. As introduced in Section~\ref{sec:CDM}, this loss is computed using the cosine distance between the projected feature embeddings from both modalities. 

After extensive evaluation and experimentation, the loss weight coefficients are set based on the relative importance of each objective and the observed training dynamics. Specifically, the segmentation loss and edge loss are weighted as $\lambda_{\text{S}} = 1.0$ and $\lambda_{\text{E}} = 0.3$, respectively. For the reconstruction loss $\lambda_{\text{R}}$, its MSE component is assigned $\lambda_{\text{mse}} = 0.2$, and the KL component is assigned $\lambda_{\text{kl}} = 0.2$. Additionally, the CDM loss is weighted by $\lambda_{\text{C}} = 0.3$ to encourage cross-modal consistency. Notably, increasing the weight $\lambda_{\text{E}}$ or $\lambda_{\text{C}}$ lead to unstable optimization gains, indicating that the current configuration effectively stabilizes multi-objective training. Additionally, the model exhibits consistent performance under minor perturbations to these weights, suggesting the robustness of the proposed weighting scheme.

All experiments are conducted on a workstation equipped with a GeForce RTX 4090 GPU and 24 GB of RAM. The optimizer used is stochastic gradient descent (SGD)~\cite{bottou2012stochastic}, with a fixed learning rate of 0.01. The training is performed for 100 epochs, with a batch size of 5 due to GPU memory limitations.

\subsection{Evaluation Metrics }

In the tree cover mapping results generated by the semantic segmentation network, there are only two classes: 1 for tree cover and 0 for non-tree. Mean IoU (mIoU) is chosen as the primary evaluation metric to assess overall segmentation performance. In addition, IoU, F1-score (F1), precision, and recall are reported specifically for the tree cover class. The calculation methods for these metrics are as follows:

\begin{align}
  \label{eq11}
\text{mIoU} &= \frac{1}{2}\sum_{c=1}^{2}\frac{\text{TP}_{c}}{\text{TP}_{c}+\text{FP}_{c}+\text{FN}_{c}},\\
\text{IoU} &= \frac{\text{TP}}{\text{TP}+\text{FP}+\text{FN}},\\
\text{Precision} &= \frac{\text{TP}}{\text{TP}+\text{FP}},\\
\text{Recall} &= \frac{\text{TP}}{\text{TP}+\text{FN}},\\
\text{F1 - score} &= 2\times \frac{\text{Precision}\times \text{Recall}}{\text{Precision} + \text{Recall}},
\end{align}
where $\text{TP}_{c}$, $\text{FP}_{c}$, and $\text{FN}_{c}$ denote the numbers of true positive, false positive, and false negative pixels for class $c$ ($c=1$ for tree cover and $c=2$ for non-tree), respectively. For class-wise metrics including IoU, Precision, Recall, and F1-score, tree cover is treated as the positive class and non-tree as the negative class. $\text{TP}$ is the number of pixels correctly predicted as tree cover (true positives), $\text{TN}$ is the number of pixels correctly predicted as non-tree (true negatives), $\text{FP}$ is the number of non-tree pixels incorrectly predicted as tree cover (false positives), $\text{FN}$ is the number of tree cover pixels incorrectly predicted as non-tree (false negatives).

\subsection{Performance Comparison}

\begin{table}[hptb]
\caption{\label{tab:experiments_zurich}
    %Some notations.
    Comparison of the evaluation metric scores(\%) among different segmentation networks of the Zurich dataset. 
    }
\resizebox{9cm}{!}{ 
\begin{tabular}{
>{\columncolor[HTML]{FFFFFF}}l |
>{\columncolor[HTML]{FFFFFF}}c |
>{\columncolor[HTML]{FFFFFF}}c 
>{\columncolor[HTML]{FFFFFF}}c 
>{\columncolor[HTML]{FFFFFF}}c 
>{\columncolor[HTML]{FFFFFF}}c }
\hline
\cellcolor[HTML]{FFFFFF}{\color[HTML]{000000} }   & \cellcolor[HTML]{FFFFFF}{\color[HTML]{000000} } & \multicolumn{4}{c}{\cellcolor[HTML]{FFFFFF}{\color[HTML]{000000} Tree Cover}}     \\ \cline{3-6} 
\multirow{-2}{*}{\cellcolor[HTML]{FFFFFF}{\color[HTML]{000000} Network}}  & \multirow{-2}{*}{\cellcolor[HTML]{FFFFFF}{\color[HTML]{000000} mIoU}} & {\color[HTML]{000000} IoU}   & {\color[HTML]{000000} F1}    & {\color[HTML]{000000} Precision}      & \multicolumn{1}{l}{\cellcolor[HTML]{FFFFFF}{\color[HTML]{000000} Recall}} \\ \hline
{\color[HTML]{000000} ResUNet~\cite{alom2018recurrent}}   & {\color[HTML]{000000} 89.61} & {\color[HTML]{000000} 86.50} & {\color[HTML]{000000} 92.33} & {\color[HTML]{000000} 90.75} & {\color[HTML]{000000} 93.97}     \\ \hline
{\color[HTML]{000000} ABCNet~\cite{li2021abcnet}}  & {\color[HTML]{000000} 90.61} & {\color[HTML]{000000} 87.46} & {\color[HTML]{000000} 93.31} & {\color[HTML]{000000} 92.35} & {\color[HTML]{000000} 94.29}     \\ \hline
{\color[HTML]{000000} PSPNet~\cite{zhao2017pyramid}}  & {\color[HTML]{000000} 90.92} & {\color[HTML]{000000} 87.90} & {\color[HTML]{000000} 93.56} & {\color[HTML]{000000} 92.08} & {\color[HTML]{000000} 95.09}     \\ \hline
{\color[HTML]{000000} CMGFNet~\cite{hosseinpour2022cmgfnet}}   & {\color[HTML]{000000} 90.79} & {\color[HTML]{000000} 87.68} & {\color[HTML]{000000} 93.44} & {\color[HTML]{000000} 92.69} & {\color[HTML]{000000} 94.19}     \\ \hline
{\color[HTML]{000000} MSSNet~\cite{dai2024robust}}  & {\color[HTML]{000000} 89.05} & {\color[HTML]{000000} 85.44} & {\color[HTML]{000000} 92.15} & {\color[HTML]{000000} 90.73} & {\color[HTML]{000000} 93.61}     \\ \hline
{\color[HTML]{000000} SegFormer~\cite{xie2021segformer}} & {\color[HTML]{000000} 89.98} & {\color[HTML]{000000} 86.62} & {\color[HTML]{000000} 92.83} & {\color[HTML]{000000} 92.18} & {\color[HTML]{000000} 93.50}     \\ \hline
{\color[HTML]{000000} GeminiFusion~\cite{jia2024geminifusion}}  & {\color[HTML]{000000} 90.35} & {\color[HTML]{000000} 87.20} & {\color[HTML]{000000} 93.16} & {\color[HTML]{000000} 91.11} & {\color[HTML]{000000} 95.31}     \\ \hline
{\color[HTML]{000000} FuseNet~\cite{hazirbas2016fusenet}}   & {\color[HTML]{000000} 91.43} & {\color[HTML]{000000} 88.50} & {\color[HTML]{000000} 93.90} & {\color[HTML]{000000} 93.65} & {\color[HTML]{000000} 94.15}     \\ \hline

{\color[HTML]{000000} ASANet~\cite{zhang2024asanet}}  & {\color[HTML]{000000} 90.05} & {\color[HTML]{000000} 86.74} & {\color[HTML]{000000} 92.90} & {\color[HTML]{000000} 91.64} & {\color[HTML]{000000} 94.19}     \\ \hline
{\color[HTML]{000000} FTransUNet~\cite{ma2024multilevel}}    & {\color[HTML]{000000} 91.54} & {\color[HTML]{000000} 88.60} & {\color[HTML]{000000} 93.96} & {\color[HTML]{000000} 94.44} & {\color[HTML]{000000} 93.48}     \\ \hline
{\color[HTML]{000000} UNetFormer~\cite{wang2022unetformer}~(ORS)}    & {\color[HTML]{000000} 88.61} & {\color[HTML]{000000} 84.93} & {\color[HTML]{000000} 91.85} & {\color[HTML]{000000} 89.59} & {\color[HTML]{000000} 94.23}     \\ \hline
{\color[HTML]{000000} UNetFormer~\cite{wang2022unetformer}}    & {\color[HTML]{000000} 92.32} & {\color[HTML]{000000} 89.71} & {\color[HTML]{000000} 94.57} & {\color[HTML]{000000} 94.03} & {\color[HTML]{000000} 95.12}     \\ \hline
{\color[HTML]{000000} \begin{tabular}[c]{@{}l@{}}Swin Transformer~\cite{liu2021swin}\\ (ORS)\end{tabular}} & {\color[HTML]{000000} 87.63} & {\color[HTML]{000000} 93.69} & {\color[HTML]{000000} 91.12} & {\color[HTML]{000000} 88.46} & {\color[HTML]{000000} 93.95}     \\ \hline
{\color[HTML]{000000} Swin Transformer~\cite{liu2021swin}}        & {\color[HTML]{000000} 91.45} & {\color[HTML]{000000} 88.48} & {\color[HTML]{000000} 93.89} & {\color[HTML]{000000} 93.65} & {\color[HTML]{000000} 94.12}     \\ \hline
{\color[HTML]{000000} \textbf{MURTreeFormer}} & {\color[HTML]{000000} \textbf{93.33}}     & {\color[HTML]{000000} \textbf{91.03}} & {\color[HTML]{000000} \textbf{95.30}} & {\color[HTML]{000000} \textbf{94.96}} & {\color[HTML]{000000} \textbf{95.64}}         \\ \hline
\end{tabular}

}
\end{table}

\begin{table}[hptb]
\caption{\label{tab:experiments_shanghai}
    %Some notations.
    Comparison of the evaluation metric scores(\%) among different segmentation networks of the Shanghai dataset. 
    }
\resizebox{9cm}{!}{ 
\begin{tabular}{
>{\columncolor[HTML]{FFFFFF}}l |
>{\columncolor[HTML]{FFFFFF}}c |
>{\columncolor[HTML]{FFFFFF}}c 
>{\columncolor[HTML]{FFFFFF}}c 
>{\columncolor[HTML]{FFFFFF}}c 
>{\columncolor[HTML]{FFFFFF}}c }
\hline
\cellcolor[HTML]{FFFFFF}{\color[HTML]{000000} }   & \cellcolor[HTML]{FFFFFF}{\color[HTML]{000000} } & \multicolumn{4}{c}{\cellcolor[HTML]{FFFFFF}{\color[HTML]{000000} Tree Cover}}     \\ \cline{3-6} 
\multirow{-2}{*}{\cellcolor[HTML]{FFFFFF}{\color[HTML]{000000} Network}}  & \multirow{-2}{*}{\cellcolor[HTML]{FFFFFF}{\color[HTML]{000000} mIoU}} & {\color[HTML]{000000} IoU}   & {\color[HTML]{000000} F1}    & {\color[HTML]{000000} Precision}      & \multicolumn{1}{l}{\cellcolor[HTML]{FFFFFF}{\color[HTML]{000000} Recall}} \\ \hline
{\color[HTML]{000000} ResUNet~\cite{alom2018recurrent}}   & {\color[HTML]{000000} 80.69} & {\color[HTML]{000000} 76.29} & {\color[HTML]{000000} 84.58} & {\color[HTML]{000000} 87.70} & {\color[HTML]{000000} 81.69}     \\ \hline
{\color[HTML]{000000} ABCNet~\cite{li2021abcnet}}  & {\color[HTML]{000000} 81.27} & {\color[HTML]{000000} 77.33} & {\color[HTML]{000000} 85.27} & {\color[HTML]{000000} 85.41} & {\color[HTML]{000000} 85.13}     \\ \hline
{\color[HTML]{000000} PSPNet~\cite{zhao2017pyramid}}  & \cellcolor[HTML]{FFFFFF}{\color[HTML]{000000} 82.89}     & {\color[HTML]{000000} 80.01} & {\color[HTML]{000000} 87.01} & {\color[HTML]{000000} 89.70} & {\color[HTML]{000000} 84.48}     \\ \hline
{\color[HTML]{000000} CMGFNet~\cite{hosseinpour2022cmgfnet}}   & {\color[HTML]{000000} 82.54} & {\color[HTML]{000000} 79.40} & {\color[HTML]{000000} 86.62} & {\color[HTML]{000000} 86.44} & {\color[HTML]{000000} 86.81}     \\ \hline
{\color[HTML]{000000} MSSNet~\cite{dai2024robust}}  & {\color[HTML]{000000} 81.14} & {\color[HTML]{000000} 76.97} & {\color[HTML]{000000} 85.04} & {\color[HTML]{000000} \textbf{90.19}} & {\color[HTML]{000000} 80.44}     \\ \hline
{\color[HTML]{000000} SegFormer~\cite{xie2021segformer}} & {\color[HTML]{000000} 83.67} & {\color[HTML]{000000} 81.42} & {\color[HTML]{000000} 87.90} & {\color[HTML]{000000} 88.86} & {\color[HTML]{000000} 86.97}     \\ \hline
{\color[HTML]{000000} GeminiFusion~\cite{jia2024geminifusion}}  & {\color[HTML]{000000} 81.17} & {\color[HTML]{000000} 77.10} & {\color[HTML]{000000} 85.12} & {\color[HTML]{000000} 87.34} & {\color[HTML]{000000} 83.02}     \\ \hline
{\color[HTML]{000000} FuseNet~\cite{hazirbas2016fusenet}}   & {\color[HTML]{000000} 84.20} & {\color[HTML]{000000} 82.22} & {\color[HTML]{000000} 88.40} & {\color[HTML]{000000} 88.22} & {\color[HTML]{000000} 88.59}     \\ \hline

{\color[HTML]{000000} ASANet~\cite{zhang2024asanet}}  & {\color[HTML]{000000} 83.72} & {\color[HTML]{000000} 81.34} & {\color[HTML]{000000} 87.85} & {\color[HTML]{000000} 87.73} & {\color[HTML]{000000} 87.98}     \\ \hline
{\color[HTML]{000000} FTransUNet~\cite{ma2024multilevel}}    & \cellcolor[HTML]{FFFFFF}{\color[HTML]{000000} 82.27}     & {\color[HTML]{000000} 79.01} & {\color[HTML]{000000} 86.37} & {\color[HTML]{000000} 88.82} & {\color[HTML]{000000} 84.05}     \\ \hline
{\color[HTML]{000000} UNetFormer~\cite{wang2022unetformer}~(ORS)}    & {\color[HTML]{000000} 79.95} & {\color[HTML]{000000} 75.19} & {\color[HTML]{000000} 83.85} & {\color[HTML]{000000} 81.66} & {\color[HTML]{000000} 86.17}     \\ \hline
{\color[HTML]{000000} UNetFormer~\cite{wang2022unetformer}}    & {\color[HTML]{000000} 83.96} & {\color[HTML]{000000} 81.92} & {\color[HTML]{000000} 88.22} & {\color[HTML]{000000} 88.27} & {\color[HTML]{000000} 88.17}     \\ \hline
{\color[HTML]{000000} \begin{tabular}[c]{@{}l@{}}Swin Transformer~\cite{liu2021swin}\\ (ORS)\end{tabular}} & \cellcolor[HTML]{FFFFFF}{\color[HTML]{000000} 79.63}     & {\color[HTML]{000000} 74.52} & {\color[HTML]{000000} 83.39} & {\color[HTML]{000000} 82.23} & {\color[HTML]{000000} 82.23}     \\ \hline
{\color[HTML]{000000} Swin Transformer~\cite{liu2021swin}}        & {\color[HTML]{000000} 82.73} & {\color[HTML]{000000} 79.80} & {\color[HTML]{000000} 86.88} & {\color[HTML]{000000} 86.13} & {\color[HTML]{000000} 87.64}     \\ \hline
{\color[HTML]{000000} \textbf{MURTreeFormer}} & {\color[HTML]{000000} \textbf{85.15}}     & {\color[HTML]{000000} \textbf{83.98}} & {\color[HTML]{000000} \textbf{89.49}} & {\color[HTML]{000000} 89.96} & {\color[HTML]{000000} \textbf{89.03}}         \\ \hline
\end{tabular}

}
\end{table}

\begin{figure*}[htbp]
  \centering
  \includegraphics[width=16cm]{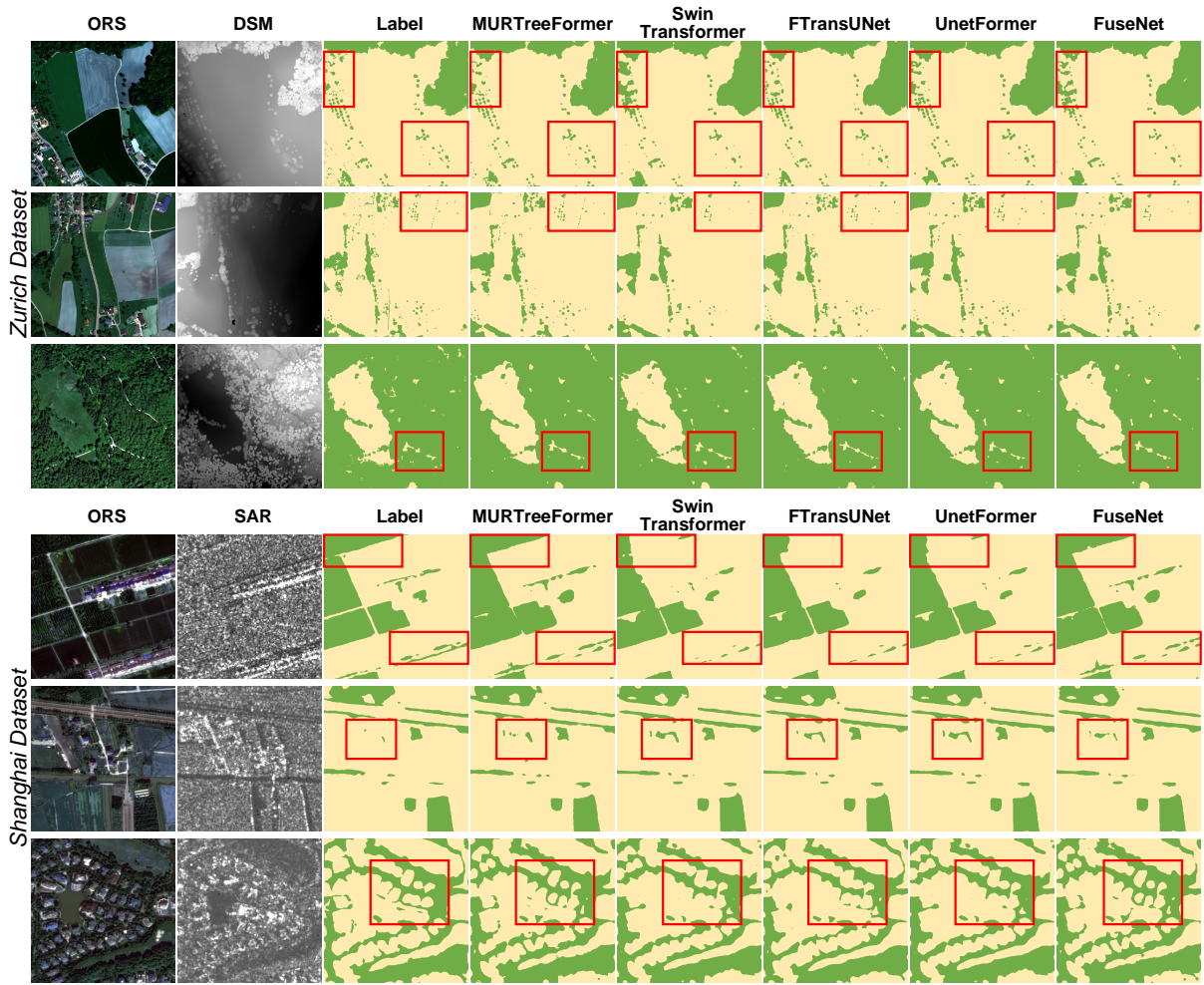}
  \caption{Segmentation results of MURTreeFormer and other networks, based on the testing dataset.}
  \label{fig:experiments}
 \end{figure*}

To further evaluate the effectiveness and generalization ability of the proposed MURTreeFormer, comprehensive comparisons are conducted with a series of state-of-the-art segmentation networks on both the Zurich and Shanghai datasets, as shown in Table~\ref{tab:experiments_zurich} and Table~\ref{tab:experiments_shanghai}. Evaluation metrics on both datasets are reported for a consistent comparison. 

The selected baseline models include both multi-modal segmentation networks (CMGFNet~\cite{hosseinpour2022cmgfnet}, GeminiFusion~\cite{jia2024geminifusion}, FuseNet~\cite{hazirbas2016fusenet}, ASANet~\cite{zhang2024asanet}, MSSNet~\cite{dai2024robust} and FTransUNet~\cite{ma2024multilevel}) and unimodal segmentation networks (ResUNet~\cite{alom2018recurrent}, ABCNet~\cite{li2021abcnet}, PSPNet~\cite{zhao2017pyramid}, SegFormer~\cite{xie2021segformer}, UNetFormer~\cite{wang2022unetformer}, and Swin Transformer~\cite{liu2021swin}). For unimodal networks, the optical and auxiliary modalities are concatenated along the channel dimension before input to the encoder. In terms of architectural paradigm, CNN-based models include ResUNet, ABCNet, PSPNet, CMGFNet, MSSNet, and FuseNet, while Transformer-based models include SegFormer, GeminiFusion, UNetFormer, ASANet, FTransUNet, and Swin Transformer. For Transformer-based encoders (e.g., Swin Transformer, GeminiFusion), the progressive decoder of MURTreeFormer without GMA and RH module is employed for fair alignment of decoding structure. To ensure fairness across all comparisons, no pretrained weights are used. All models are trained with the same batch size of 5. Each network is independently trained three times, and the average of the segmentation metrics is reported to ensure robustness and stability.

\begin{table*}[hptb]
\caption{\label{tab:ablation_zurich}
    %Some notations.
    Comparison of the evaluation metric scores(\%) for different modules of the Zurich dataset. 
    }
\centering
\begin{tabular}{
>{\columncolor[HTML]{FFFFFF}}c |ccccc|cccc}
\hline
\cellcolor[HTML]{FFFFFF}{\color[HTML]{000000} }   & \multicolumn{1}{c|}{}    & \multicolumn{2}{c|}{\cellcolor[HTML]{FFFFFF}{\color[HTML]{000000} Encoder}}    & \multicolumn{2}{c|}{Decoder}  & \multicolumn{4}{c}{\cellcolor[HTML]{FFFFFF}{\color[HTML]{000000} Tree Cover}} \\ \cline{3-10} 
\multirow{-2}{*}{\cellcolor[HTML]{FFFFFF}{\color[HTML]{000000} Network}} & \multicolumn{1}{c|}{\multirow{-2}{*}{Backbone}} & \cellcolor[HTML]{FFFFFF}{\color[HTML]{000000} SURM} & \multicolumn{1}{c|}{CDM} & \multicolumn{1}{c|}{GMA} & RH & \cellcolor[HTML]{FFFFFF}{\color[HTML]{000000} IoU} & \cellcolor[HTML]{FFFFFF}{\color[HTML]{000000} F1} & \cellcolor[HTML]{FFFFFF}{\color[HTML]{000000} Precision} & \cellcolor[HTML]{FFFFFF}{\color[HTML]{000000} Acc}   \\ \hline
\cellcolor[HTML]{FFFFFF}{\color[HTML]{000000} R1} & o & \cellcolor[HTML]{FFFFFF}{\color[HTML]{000000} }     &   &   &    & \cellcolor[HTML]{FFFFFF}{\color[HTML]{000000} 88.48} & \cellcolor[HTML]{FFFFFF}{\color[HTML]{000000} 93.89} & \cellcolor[HTML]{FFFFFF}{\color[HTML]{000000} 93.65} & \cellcolor[HTML]{FFFFFF}{\color[HTML]{000000} 94.12} \\ \hline
R2  & o & o     &   &   &    & 90.27     & 94.89     & 93.73     & 96.07  \\ \hline
R3  & o & & o &   &    & 90.03     & 94.75     & 94.16     & 95.36  \\ \hline
R4  & o & &   & o &    & 89.70     & 94.57     & 94.03     & 95.12  \\ \hline
R5  & o & &   &   & o  & 89.48     & 94.45     & 94.40     & 94.50  \\ \hline
{\color[HTML]{000000} R6}  & o & \cellcolor[HTML]{FFFFFF}{\color[HTML]{000000} }     & o & o & o  & \cellcolor[HTML]{FFFFFF}{\color[HTML]{000000} 90.42} & \cellcolor[HTML]{FFFFFF}{\color[HTML]{000000} 94.97} & \cellcolor[HTML]{FFFFFF}{\color[HTML]{000000} 93.87} & \cellcolor[HTML]{FFFFFF}{\color[HTML]{000000} \textbf{96.10}} \\ \hline
{\color[HTML]{000000} R7}  & o & \cellcolor[HTML]{FFFFFF}{\color[HTML]{000000} o}    &   & o & o  & \cellcolor[HTML]{FFFFFF}{\color[HTML]{000000} 90.90} & \cellcolor[HTML]{FFFFFF}{\color[HTML]{000000} 95.23} & \cellcolor[HTML]{FFFFFF}{\color[HTML]{000000} 94.45} & \cellcolor[HTML]{FFFFFF}{\color[HTML]{000000} 96.03} \\ \hline
\cellcolor[HTML]{FFFFFF}{\color[HTML]{000000} R8} & o & \cellcolor[HTML]{FFFFFF}{\color[HTML]{000000} o}    & o &   & o  & \cellcolor[HTML]{FFFFFF}{\color[HTML]{000000} 90.77} & \cellcolor[HTML]{FFFFFF}{\color[HTML]{000000} 95.16} & \cellcolor[HTML]{FFFFFF}{\color[HTML]{000000} 94.84} & \cellcolor[HTML]{FFFFFF}{\color[HTML]{000000} 95.49} \\ \hline
\cellcolor[HTML]{FFFFFF}{\color[HTML]{000000} MURTreeFormer} & o & \cellcolor[HTML]{FFFFFF}{\color[HTML]{000000} o}    & o & o & o  & \cellcolor[HTML]{FFFFFF}{\color[HTML]{000000} \textbf{91.03}} & \cellcolor[HTML]{FFFFFF}{\color[HTML]{000000} \textbf{95.30}} & \cellcolor[HTML]{FFFFFF}{\color[HTML]{000000} \textbf{94.96}} & \cellcolor[HTML]{FFFFFF}{\color[HTML]{000000} 95.64} \\ \hline
\end{tabular}

\vspace{5mm}

\caption{\label{tab:ablation_shanghai}
    %Some notations.
    Comparison of the evaluation metric scores(\%) for different modules of the Shanghai dataset. 
    }
\centering 
\begin{tabular}{
>{\columncolor[HTML]{FFFFFF}}c |ccccc|cccc}
\hline
\cellcolor[HTML]{FFFFFF}{\color[HTML]{000000} }   & \multicolumn{1}{c|}{}    & \multicolumn{2}{c|}{\cellcolor[HTML]{FFFFFF}{\color[HTML]{000000} Encoder}}    & \multicolumn{2}{c|}{Decoder}  & \multicolumn{4}{c}{\cellcolor[HTML]{FFFFFF}{\color[HTML]{000000} Tree Cover}}    \\ \cline{3-10} 
\multirow{-2}{*}{\cellcolor[HTML]{FFFFFF}{\color[HTML]{000000} Network}} & \multicolumn{1}{c|}{\multirow{-2}{*}{Backbone}} & \cellcolor[HTML]{FFFFFF}{\color[HTML]{000000} SURM} & \multicolumn{1}{c|}{CDM} & \multicolumn{1}{c|}{GMA} & RH & \cellcolor[HTML]{FFFFFF}{\color[HTML]{000000} IoU} & \cellcolor[HTML]{FFFFFF}{\color[HTML]{000000} F1} & \cellcolor[HTML]{FFFFFF}{\color[HTML]{000000} Precision} & \cellcolor[HTML]{FFFFFF}{\color[HTML]{000000} Recall}   \\ \hline
\cellcolor[HTML]{FFFFFF}{\color[HTML]{000000} R1} & o & \cellcolor[HTML]{FFFFFF}{\color[HTML]{000000} }     &   &   &    & \cellcolor[HTML]{FFFFFF}{\color[HTML]{000000} 79.80} & \cellcolor[HTML]{FFFFFF}{\color[HTML]{000000} 86.88} & \cellcolor[HTML]{FFFFFF}{\color[HTML]{000000} 86.13} & \cellcolor[HTML]{FFFFFF}{\color[HTML]{000000} 87.64} \\ \hline
R2  & o & o     &   &   &    & 82.40     & 88.52     & 90.45     & 86.67     \\ \hline
R3  & o & & o &   &    & 83.03     & 88.91     & 89.89     & 87.95     \\ \hline
R4  & o & &   & o &    & 81.79     & 88.14     & 90.06     & 86.29     \\ \hline
R5  & o & &   &   & o  & 81.42     & 87.90     & 88.86     & 86.98     \\ \hline
{\color[HTML]{000000} R6}  & o & \cellcolor[HTML]{FFFFFF}{\color[HTML]{000000} }     & o & o & o  & \cellcolor[HTML]{FFFFFF}{\color[HTML]{000000} 83.56} & \cellcolor[HTML]{FFFFFF}{\color[HTML]{000000} 89.23} & \cellcolor[HTML]{FFFFFF}{\color[HTML]{000000} 90.09} & \cellcolor[HTML]{FFFFFF}{\color[HTML]{000000} 88.39} \\ \hline
{\color[HTML]{000000} R7}  & o & \cellcolor[HTML]{FFFFFF}{\color[HTML]{000000} o}    &   & o & o  & \cellcolor[HTML]{FFFFFF}{\color[HTML]{000000} 81.58} & \cellcolor[HTML]{FFFFFF}{\color[HTML]{000000} 88.00} & \cellcolor[HTML]{FFFFFF}{\color[HTML]{000000} 89.65} & \cellcolor[HTML]{FFFFFF}{\color[HTML]{000000} 86.41} \\ \hline
\cellcolor[HTML]{FFFFFF}{\color[HTML]{000000} R8} & o & \cellcolor[HTML]{FFFFFF}{\color[HTML]{000000} o}    & o &   & o  & \cellcolor[HTML]{FFFFFF}{\color[HTML]{000000} 83.79} & \cellcolor[HTML]{FFFFFF}{\color[HTML]{000000} 89.38} & \cellcolor[HTML]{FFFFFF}{\color[HTML]{000000} \textbf{90.97}} & \cellcolor[HTML]{FFFFFF}{\color[HTML]{000000} 87.84} \\ \hline
\cellcolor[HTML]{FFFFFF}{\color[HTML]{000000} MURTreeFormer} & o & \cellcolor[HTML]{FFFFFF}{\color[HTML]{000000} o}    & o & o & o  & \cellcolor[HTML]{FFFFFF}{\color[HTML]{000000} \textbf{83.98}} & \cellcolor[HTML]{FFFFFF}{\color[HTML]{000000} \textbf{89.49}} & \cellcolor[HTML]{FFFFFF}{\color[HTML]{000000} 89.96} & \cellcolor[HTML]{FFFFFF}{\color[HTML]{000000} \textbf{89.03}} \\ \hline
\end{tabular}

\end{table*}

On the Zurich dataset, MURTreeFormer achieves the highest mIoU of 93.33\%, significantly outperforming UNetFormer (92.32\%) and FTransUNet (91.54\%), demonstrating its excellent capability in large-scale tree cover mapping. Moreover, it obtains the best performance in tree cover IoU (91.03\%) and F1 score (95.30\%), indicating a stronger ability to extract fine-grained vegetation structures. On the Shanghai dataset, MURTreeFormer also delivers the best results, reaching the highest mIoU of 85.15\%, surpassing competitive methods such as SegFormer (83.67\%) and UNetFormer (83.96\%). It also achieves the top scores in tree cover IoU (83.98\%) and F1 score (89.49\%), confirming its robustness and generalization ability in complex and large-scale urban environments. Compared with other Transformer-based networks (e.g., ASANet and FTransUNet), MURTreeFormer shows consistent improvements across various metrics, which validates the effectiveness of its encoder-decoder design and modality-aware fusion modules. Similarly, when compared to CNN-based architectures such as ResUNet, PSPNet, or CMGFNet, Transformer-based models generally achieve better performance, especially in terms of IoU and F1 scores, due to their superior ability in modeling long-range dependencies and capturing structural semantics. Furthermore, unimodal and multimodal segmentation strategies were compared. For unimodal baselines, ORS and auxiliary modalities were concatenated as input (e.g., in Swin Transformer and UNetFormer), while multimodal networks explicitly integrate modality-specific branches and fusion mechanisms. The proposed method shows clear advantages over these unimodal approaches, especially on the Shanghai dataset, where the spatial resolution gap between modalities (0.68~m ORS vs. 3~m SAR) poses additional challenges. The consistent performance gain confirms that leveraging complementary information from different modalities significantly enhances tree cover segmentation quality. Additionally, when feeding only ORS images into UNetFormer and Swin Transformer, the results are consistently inferior to their multimodal counterparts, further confirming that fusing complementary information from different modalities significantly enhances tree cover segmentation performance.

The visual comparisons further corroborate these findings. As illustrated in Fig.~\ref{fig:experiments}, MURTreeFormer demonstrates superior capability in capturing fragmented and densely distributed tree regions, whereas other methods tend to miss small patches or produce blurred boundaries. In particular, on the Zurich dataset, MURTreeFormer excels in delineating individual tree crowns, clearly distinguishing them from the background and neighboring trees. This aligns well with the design goal of the proposed network—to recover high-resolution vegetation details through effective multi-modal feature modeling and uncertainty-aware refinement.

\subsection{Ablation Experiments}

To comprehensively assess the contributions of each proposed component in MURTreeFormer, a series of ablation experiments are conducted on both the Zurich and Shanghai datasets. Each module, including the selective uncertainty-guided reconstruction module (SURM), the cross-modal distillation module (CDM), the gradient magnitude attention (GMA), and the shallow feature refinement head (RH), is incrementally incorporated into a Swin Transformer with progressive decoder baseline, and its impact is evaluated through both quantitative metrics and qualitative visualizations.

The results are summarized in Table~\ref{tab:ablation_zurich} and Table~\ref{tab:ablation_shanghai}. From R2 to R5, the inclusion of each individual module consistently improves the performance over the baseline. Among them, CDM yields the most significant gain, indicating that aligning heterogeneous features across modalities allows the model to better integrate tree cover semantics. In contrast, RH shows the least improvement in quantitative metrics, which may be attributed to the fact that RH is particularly effective at refining fine-grained structures of tree crowns, leading to visually improved segmentation, while these refinements contribute less to global accuracy metrics.

From R6 to R8, different components (SURM, CDM, and GMA) are selectively removed from the full model to examine their relative importance. It can be observed that both SURM and CDM play essential roles at the encoder side. Notably, SURM exhibited stronger improvements on the Shanghai dataset, while CDM is more beneficial on Zurich. This suggests that CDM is particularly effective when aligning optical and DSM features, whereas SURM is better suited for reconstructing optical–SAR discrepancies under high uncertainty.

The complete model, MURTreeFormer, which incorporates all four modules, achieves the highest performance with 91.03\% IoU, 95.30\% F1, and 94.96\% precision on Zurich, and 83.98\% IoU, 89.49\% F1, and 89.03\% recall on Shanghai. These results highlight the synergistic benefits of the encoder and decoder designs in improving tree cover segmentation.

\begin{figure}[htbp]
  \centering
  \includegraphics[width=9cm]{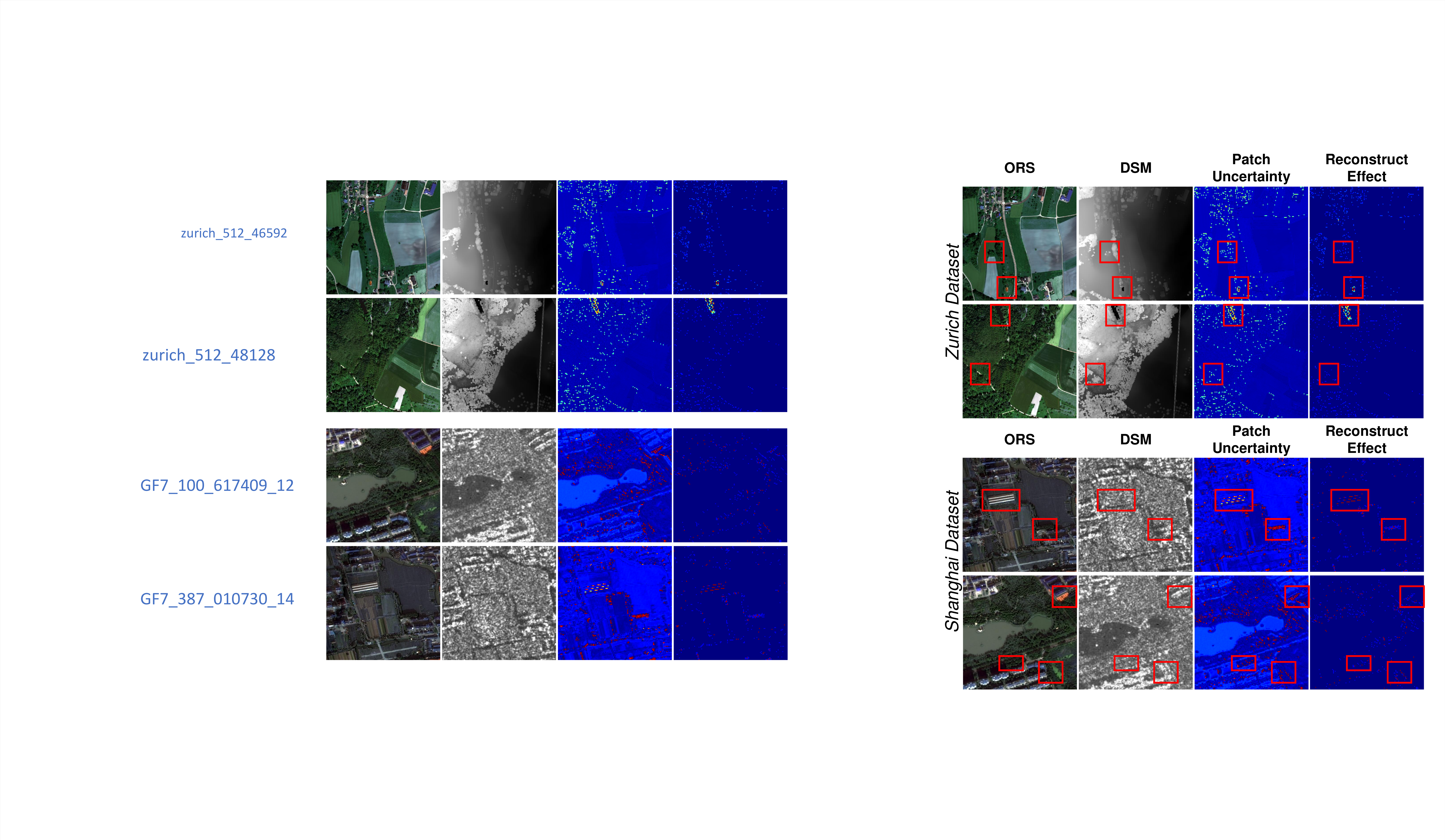}
  \caption{Visualization of patch-level uncertainty estimation and reconstruction effects on the Zurich and Shanghai datasets. In the “Patch Uncertainty” column, the top 500 most uncertain patches (with the highest scores in the uncertainty map) are highlighted to distinguish them from the remaining regions. In the “Reconstruct Effect” column, the patch-wise MSE between the reconstructed and original auxiliary modality features is visualized as a heatmap. Lower brightness indicates a smaller discrepancy between the reconstructed patch (derived from the primary modality) and the original auxiliary input, suggesting more consistent cross-modal semantics.
}
\label{fig:ablation_SURM}
\end{figure}

To further validate the effectiveness of each module, qualitative visualizations are conducted. As shown in Fig.~\ref{fig:ablation_SURM}, the contribution of SURM is demonstrated. The “patch uncertainty” column highlights the top 500 most uncertain patches, selected according to the method described in Section~\ref{sec:SURM}. These patches often correspond to regions affected by modality discrepancy, environmental variation, or imaging artifacts. The “reconstruction effect” column shows the patch-wise MSE values between the reconstructed features and the original auxiliary modality features. Since the reconstructed features are generated from the primary modality's latent distribution, a lower MSE indicates that the primary modality provides consistent and reliable semantic cues for reconstruction, while a higher value implies modality inconsistency or ambiguity. It can be observed that after SURM-based reconstruction, the uncertainty in tree covered regions is substantially reduced, with visually improved semantic structures. This suggests that SURM selectively enhances semantically important but uncertain regions, generating higher-quality auxiliary features that facilitate subsequent cross-modal fusion and tree cover segmentation.

\begin{figure}[htbp]
\centering
\includegraphics[width=8cm]{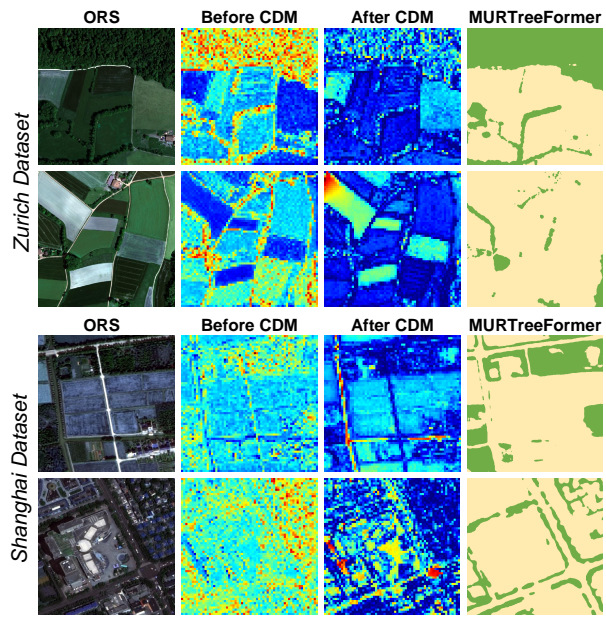}
\caption{
Visualization of the proposed Cross-Modal Distillation Module (CDM). The heatmaps depict the patch-wise feature discrepancy (measured by mean squared error) between the primary and auxiliary modalities, before and after applying CDM. A noticeable reduction in discrepancy within tree covered regions after CDM indicates enhanced feature alignment, contributing to improved segmentation performance of MURTreeFormer.
}
\label{fig:ablation_CDM}
\end{figure}

Fig.~\ref{fig:ablation_CDM} illustrates the effectiveness of CDM in aligning cross-modal features. The patch-wise MSE between projected optical and auxiliary modality features is visualized before and after CDM is applied. Prior to alignment, tree covered regions show large modality discrepancies. After CDM training, these discrepancies are significantly reduced, particularly in vegetated and structured areas. This indicates that CDM adaptively enhances alignment in semantically important regions, improving cross-modal consistency and ultimately supporting better tree cover segmentation. 

\begin{figure}[htbp]
  \centering
  \includegraphics[width=8cm]{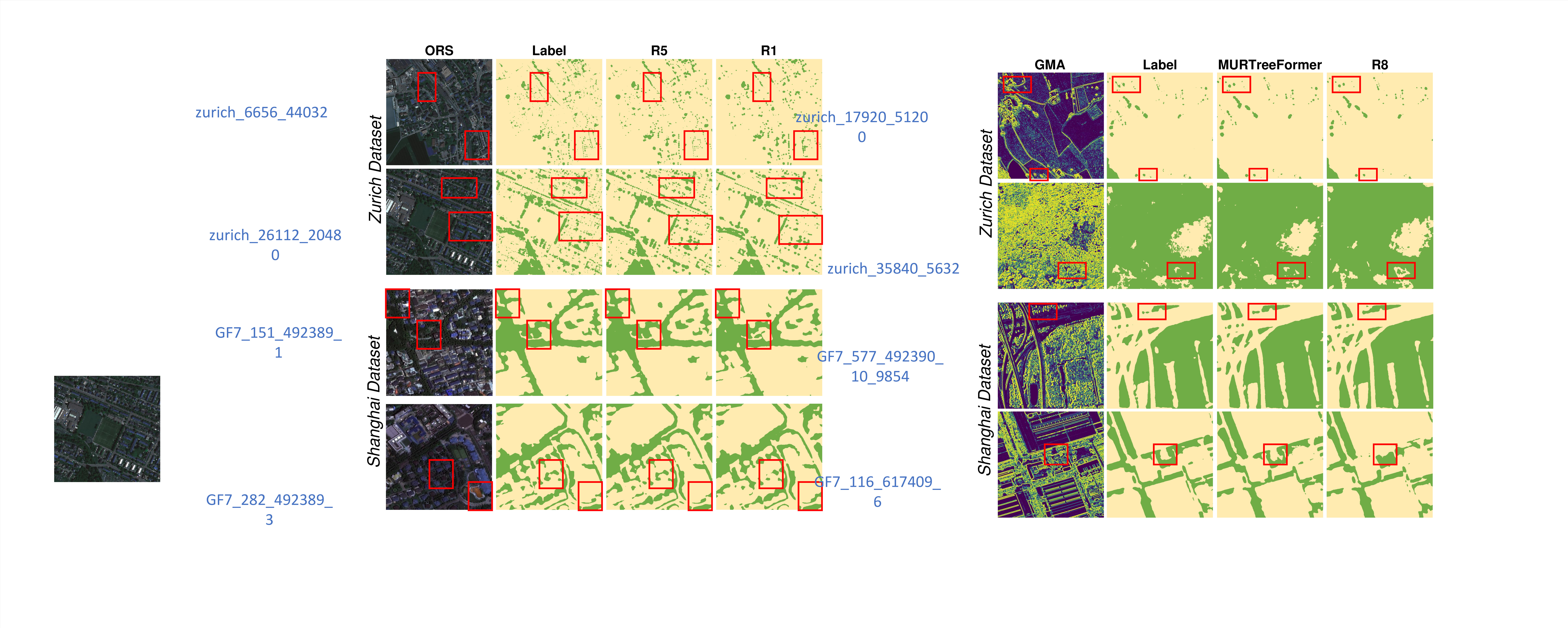}
  \caption{
    Visualization of the proposed Gradient Magnitude Attention (GMA). In this task, the luminance map of the primary modality (ORS) is used to compute the attention map, as shown in the GMA column. Compared to the R8, the inclusion of GMA in MURTreeFormer improves tree cover delineation, particularly at fine-grained boundaries.
  }
  \label{fig:ablation_GMA}
\end{figure}

In Fig.~\ref{fig:ablation_GMA}, the attention maps computed based on luminance gradients are visualized to demonstrate the effectiveness of GMA for tree cover segmentation. In both Zurich and Shanghai datasets, the ORS modality is used as the primary input. The resulting attention maps guide the decoder to focus on texture-continuous and edge-salient regions, allowing the network to better distinguish tree cover from spectrally similar vegetation, even under subtle appearance variations. Compared with the R8 configuration (without GMA), MURTreeFormer exhibits enhanced sensitivity to tree cover structures across multiple modalities, especially when dealing with high-resolution inputs. As shown in Table~\ref{tab:ablation_zurich} and Table~\ref{tab:ablation_shanghai} (e.g., comparisons between R1, R4, and R8), the inclusion of GMA consistently improves the decoder’s ability to represent and refine tree cover features. This further confirms that gradient-guided attention can effectively enhance the decoder’s sensitivity to textural continuity and contour sharpness of tree crowns, leading to more coherent segmentation outputs. While the improvements brought by GMA are more prominent in complex or texture-rich regions, the enhancements may be less noticeable in homogeneous areas where tree cover boundaries are already well defined.

\begin{figure}[htbp]
  \centering
  \includegraphics[width=8cm]{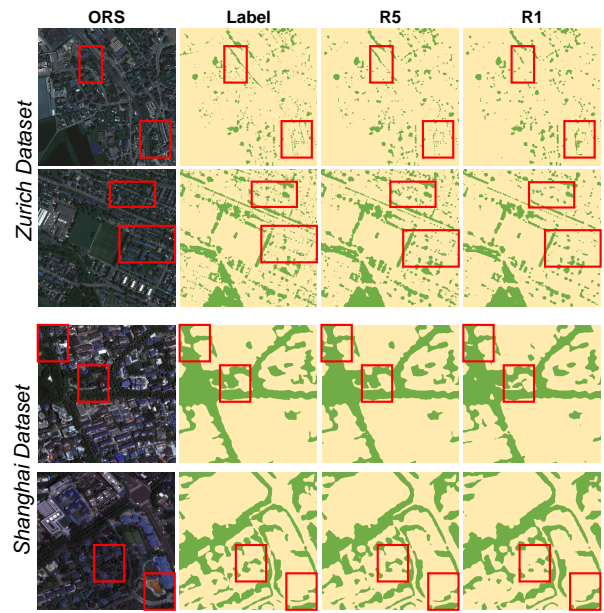}
  \caption{Visual comparison of tree cover segmentation results with and without the proposed Refinement Head (RH). Compared to the baseline output (R1), the inclusion of RH in R5 results in sharper and more complete delineation of tree crowns, especially in fragmented or sparse vegetation regions.}
  \label{fig:ablation_RH}
 \end{figure}

Additionally, the benefit of the Refinement Head (RH) is demonstrated in Fig.~\ref{fig:ablation_RH}. Although the numerical performance gain is modest, RH effectively enhances the delineation of fine-grained tree structures. Comparing R5 with R1, where RH is the only added component, reveals visibly improved boundary sharpness and better separation of sparse or fragmented tree crowns, particularly in complex urban interspaces. The RH module adopts a lightweight two-step upsampling strategy, and this progressive recovery allows the network to integrate richer spatial features from intermediate stages, aiding in the reconstruction of detailed boundaries. Despite its effectiveness, RH introduces only a small number of convolutional parameters and negligible additional FLOPs (less than 2\% of the overall backbone), ensuring efficient inference. These results confirm the RH’s complementary role in refining local spatial details, compensating for the global representations captured by the main decoder path.

%% file: 5Conclusion.tex
This paper investigates the often-overlooked aleatoric uncertainty problem in multi-modal semantic segmentation of remote sensing imagery. To effectively detect and mitigate patch-level uncertainty, we propose MURTreeFormer, a novel multi-modal semantic segmentation framework specifically designed for fine-grained tree cover interpretation. To address the challenges posed by stochastic uncertainty and structural ambiguity in tree coverage area, four dedicated modules are introduced: the SURM module for patch-level uncertainty estimation and reconstruction, the CDM module for cross-modal feature alignment, the GMA module for luminance-aware attention guidance, and the RH module for detail-preserving refinement.

Extensive experiments on Zurich and Shanghai datasets demonstrate that each of the proposed components contributes positively to overall segmentation performance. The whole MURTreeFormer model consistently outperforms strong baselines and state-of-the-art alternatives, achieving new benchmarks in tree cover segmentation.

Future work will aim to generalize the proposed framework to wider geo-environmental and ecological applications, including multi-class semantic segmentation and interpretable change detection.